%% file: SAMIGrad.tex
\documentclass[useAMS,usenatbib,usegraphicx]{mn2e}
\usepackage{amsmath}
\usepackage{pifont}
\usepackage{color}
\usepackage{multirow}

\newif\ifAMStwofonts
\AMStwofontstrue

\newcommand{\simlt}{\lower.5ex\hbox{$\; \buildrel < \over \sim \;$}}
\newcommand{\simgt}{\lower.5ex\hbox{$\; \buildrel > \over \sim \;$}}
\newcommand{\kms}{\rm km\,s$^{-1}$}

\newcommand{\mgfe}{$\rm [Mg/Fe]$}
\newcommand{\afe}{$\rm [\alpha /Fe]$}
\newcommand{\cfe}{$\rm [C/Fe]$}

\newcommand{\mgb}{$\rm Mgb5170$}
\newcommand{\fei}{$\rm Fe4383$}
\newcommand{\feii}{$\rm Fe5270$}
\newcommand{\feiii}{$\rm Fe5335$}
\newcommand{\cfs}{$\rm C4668$}
\newcommand{\hbo}{$\rm H\beta_o$}
\newcommand{\hgF}{$\rm H\gamma_F$}
\newcommand{\zh}{$\rm [Z/H]$}

\title[SAMI: Population gradients in ETGs]
{The SAMI Galaxy Survey: Stellar population radial gradients in early-type galaxies}
\author[I. Ferreras et al.]
{Ignacio Ferreras$^{1,2,3}$\thanks{E-mail: iferreras@iac.es}, 
Nicholas Scott$^{4,5}$, Francesco La Barbera$^6$, Scott M. Croom$^{4,5}$,\and
Jesse van de Sande$^{4,5}$, Andrew Hopkins$^{7}$, Matthew Colless$^{5,8}$,
Tania Barone$^{4,5,8}$, \and
Francesco d'Eugenio$^{5,8,9}$, Joss Bland-Hawthorn$^{4,5}$, Sarah Brough$^{5,10}$,\and
Julia J. Bryant$^{4,5,11}$, Iraklis S. Konstantopoulos$^{12}$, Claudia Lagos$^{13}$,
Jon S. Lawrence$^{7}$, \and
Angel L\'opez-S\'anchez$^{7}$, Anne M. Medling$^{8,13}$, Matt S. Owers$^{14,15}$,
Samuel N. Richards$^{16}$
\medskip\\
$^1$ Mullard Space Science Laboratory, University College London, 
Holmbury St Mary, Dorking, Surrey RH5 6NT, UK\\
$^2$ Instituto de Astrof{\'i}sica de Canarias, Calle V{\'i}a L{\'a}ctea s/n,
E38205, La Laguna, Tenerife, Spain\\
$^3$ Departamento de Astrof{\'i}sica, Universidad de La Laguna (ULL), E-38206 La Laguna,
Tenerife, Spain\\
$^4$ Sydney Institute for Astronomy, School of Physics, University of Sydney, NSW 2006, Australia\\
$^5$ ARC Centre of Excellence for All Sky Astrophysics in 3~Dimensions (ASTRO 3D)\\
$^6$ INAF-Osservatorio Astronomico di Capodimonte, sal. Moiariello 16, I-80131 Napoli, Italy\\
$^7$ Australian Astronomical Optics, Macquarie University, 105 Delhi Rd, North Ryde, NSW 2113, Australia\\
$^8$ Research School of Astronomy and Astrophysics, Australian National University, Canberra, ACT 2611, Australia\\
$^9$ Sterrenkundig Observatorium, Universiteit Gent, Krijgslaan 281 S9, B-9000 Gent, Belgium\\
$^{10}$ School of Physics, University of New South Wales, NSW 2052, Australia\\
$^{11}$ Australian Astronomical Optics, AAO-USydney, School of Physics, University of Sydney, NSW 2006, Australia\\
$^{12}$ Atlassian 341 George St Sydney, NSW 2000, Australia\\
$^{13}$ ICRAR, M468, University of Western Australia, 35 Stirling Hwy, Crawley, WA 6009, Australia\\
$^{13}$ Ritter Astrophysical Research Center, University of Toledo, Toledo, OH 43606, USA\\
$^{14}$ Department of Physics and Astronomy, Macquarie University, NSW 2109, Australia\\
$^{15}$ Astronomy, Astrophysics and Astrophotonics Research Centre, Macquarie University, Sydney, NSW 2109, Australia\\
$^{16}$ SOFIA, USRA, NASA Ames Research Center, Building N232, M/S 232-12,
P.O. Box 1, Moffett Field, CA 94035-0001, USA
}

\begin{document}
\date{Submitted to MNRAS (arXiv version)}
\pagerange{\pageref{firstpage}--\pageref{lastpage}} \pubyear{2019}
\maketitle
\label{firstpage}


\begin{abstract}
We study the internal radial gradients of the stellar populations in a
sample comprising 522 early-type galaxies (ETGs) from the SAMI
(Sydney- AAO Multi-object Integral field spectrograph) Galaxy 
Survey.  We stack the spectra of individual spaxels in radial bins,
and derive basic stellar population properties: total metallicity ([Z/H]), [Mg/Fe],
[C/Fe] and age.  The radial gradient ($\nabla$) and central value of
the fits (evaluated at R$_e$/4) are compared against a set of six
possible drivers of the trends. We find that velocity dispersion ($\sigma$) --
or, equivalently gravitational potential -- is the dominant driver of
the chemical composition gradients. Surface mass density is also
correlated with the trends, especially with stellar age. The decrease
of $\nabla$[Mg/Fe] with increasing $\sigma$ is contrasted by a rather
shallow dependence of $\nabla$[Z/H] with $\sigma$ (although this
radial gradient is overall rather steep). This result, along with a shallow age
slope at the massive end, imposes stringent constraints on the
progenitors of the populations that contribute to the formation of the
outer envelopes of ETGs. The SAMI sample is split between a `field'
sample and a cluster sample. Only weak environment-related
differences are found, most notably a stronger dependence of central total
metallicity ([Z/H]$_{e4}$) with $\sigma$, along with a marginal
trend of $\nabla$[Z/H] to steepen in cluster galaxies, a result that is not
followed by [Mg/Fe]. The results presented here serve as constraints
on numerical models of the formation and evolution of ETGs.
\end{abstract} 


\section{Introduction}
\label{Sec:Intro}

Radial gradients of the chemical composition of the stellar
populations of early-type galaxies (ETGs) encode valuable information
about their build-up process \citep{Larson:74}. At present, the
2-stage formation scenario \citep[e.g.][]{Oser:10} constitutes a
simplified yet insightful description of galaxy formation, especially
at the massive end. In this framework, the stellar content of galaxies is split into an
in-situ component, typically formed during the early collapse of the
gas in the fledgling halo, followed by subsequent merging events in
which stars, previously formed ex-situ, are supplied by infalling
satellite galaxies. Subsequent in-situ formation is also
possible via accretion and cooling of gas. In massive ETGs this
separation allows us to propose a simplified scenario consisting of an
early and intense phase during which a massive core is formed, along with a 
later accretion phase contributed by mergers. The stellar populations of
massive ETGs are mostly old, enabling us to cleanly split their formation
history into a core, formed in-situ at early times, and an envelope,
produced by the later, ex-situ, phase.

The presence of massive, nearly quiescent cores at high redshift
\citep[z$\sim$2--3, see, e.g.,][]{Daddi:05,Trujillo:06,vdK:08}
suggests that a single in-situ phase is not capable of producing the
massive ETGs we see today, and radial variations within ETGs can be
exploited to understand the role of the ex-situ
phase \citep[e.g.,][]{Lackner:12,Hirschmann:15}.  Moreover, variations
between field and cluster environments are expected since the latter
represents an ``accelerated'' version of the former, as higher density
regions collapse earlier.

This paper looks for clues in the formation of ETGs via
intrinsic radial gradients of the underlying stellar populations.
The advent of surveys based on  Integral Field
Spectroscopy has transformed the field of galaxy evolution
\citep[SAURON, ATLAS$^{\rm 3D}$, SAMI, CALIFA, MaNGA,][]{Bacon:01,Cap:11,Croom:12,SFS:12,Bundy:15}
 enabling spatially resolved studies of all information
accessible to spectroscopy. Vast amounts of information are encoded into the
datacubes that are now routinely studied to explore the dynamical
state and chemical properties of the stellar and gaseous phases of
galaxies.

Differences between the in-situ and ex-situ components will be present
not only in the stellar kinematics at large radii, but also in the
stellar population properties. The longer dynamical timescales in the
outer envelopes of galaxies imply that these regions fare better at
preserving information related to the past merger history.  Radial
gradients in age and chemical composition reveal variations in the
star formation histories \citep[see, e.g.,][]{Greene:15,RGD:15}, 
including properties such as the stellar initial mass function
\citep[IMF, e.g.,][]{XSH}, that reflect a
fundamentally different mode of star formation during the early
in-situ phase.

Simulations also reveal important signatures in population gradients.
At large radii, the stellar content appears to be predominantly driven
by the accretion of incoming satellite galaxies. Computer models
of galaxy formation that include feedback
prescriptions show that a substantial contribution from winds is
needed to account for the steep metallicity gradients
observed \citep{Hirschmann:15}.  The Illustris ETGs feature steeper
metallicity profiles when their mass assembly history is less
extended \citep{Cook:16}. In principle, the steepest gradients should
be expected in an orderly monolithic collapse, while mergers would
act towards washing out these gradients.

\begin{figure}
\begin{center}
\includegraphics[width=85mm]{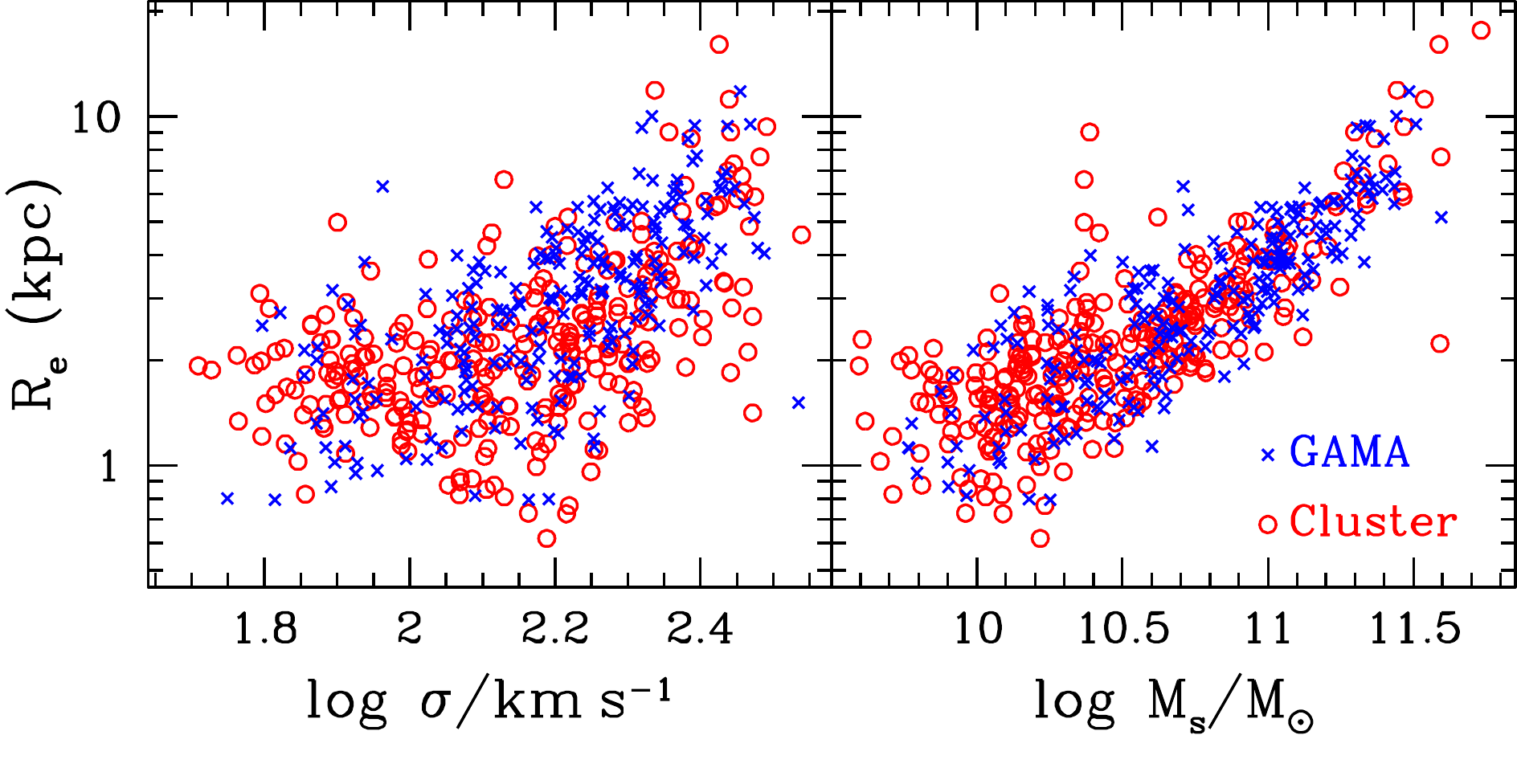}
\end{center}
\caption{Sample of SAMI early-type galaxies studied in this paper,
showing the effective radius (R$_e$) against velocity dispersion
($\sigma$, left) and stellar mass (M$_s$, right), both shown on
a logarithmic scale. The sample is split
between GAMA-selected galaxies (blue crosses, representing field and
group ETGs) and cluster galaxies (red circles).
\label{fig:Sample1}}
\end{figure}

The population gradients measured in a volume-limited sample of 95
massive early-type galaxies from the MASSIVE survey found a strong
trend of stellar age and [Mg/Fe] with velocity dispersion, contrasting
with a weaker correlation when stellar mass is
considered \citep{Greene:15}. This behaviour would suggest that
galaxies with high velocity dispersion are more efficient at
transforming gas into stars. However, this trend disappears at larger
radii ($\sim$1--1.5\,R$_e$), suggesting a complex contribution of
stellar populations in the outer regions, as they are formed {\sl
ex-situ} from a range of merging satellites.
\citet{Boardman:17} found that the kinematics in a sample of 12 ETGs,
detected in H{\sc I}, did not show any variations out to three
effective radii, supporting the idea that these galaxies have not
undergone dry major merging at late times.  However, the IFU data
revealed substantial population gradients consistent with some level
of interaction in recent times. More recently, \citet{IMN:18} explored
a sample of 45 ETGs from the CALIFA IFU survey, finding significant
radial gradients of metallicity, that increase with velocity
dispersion. In contrast, no gradient was detected with respect to
[Mg/Fe] (also note \citealt{PSB:14} for a comparative study in disks).

\begin{figure}
\begin{center}
\includegraphics[width=85mm]{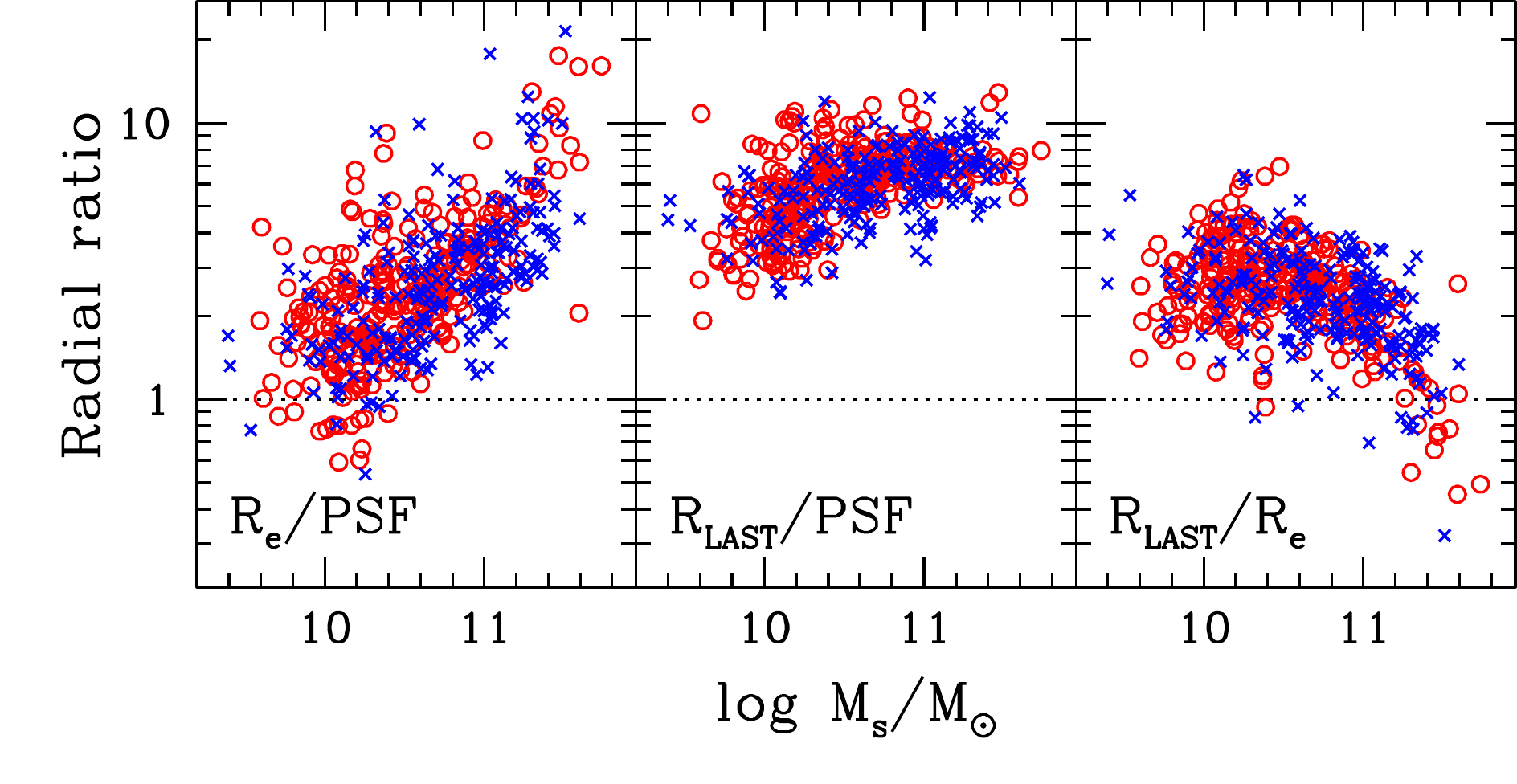}
\end{center}
\caption{Distribution of ETG sample sizes, showing 
from left to right, the effective radius and outermost radial bin
-- in units of the Point Spread Function (PSF) -- and the 
ratio of the two. The radial extent of the PSF 
is taken at the Half Width at Half-Maximum, individually for each
galaxy.  The sample is split between GAMA-selected galaxies (blue
crosses, representing field and group ETGs) and cluster galaxies (red
circles).
\label{fig:Sample2}}
\end{figure}

\citet{Goddard:17} presented an analysis of population
gradients in a large sample of 721 galaxies from the SDSS-IV MaNGA
survey, covering a wide range of stellar mass and morphology. Their
results concerning the subsample of early-type galaxies reveal small age
gradients, and negative metallicity gradients, without any significant
correlation with galaxy environment (various definitions of
environment were presented, namely nearest neighbours, gravitational
tidal strength and a central/satellite split). Such a result is at
odds with the trends presented in the SDSS-based Spider sample of
ETGs, which suggested significantly younger ages in centrals
\citep{FLB:14}, and variation in population radial gradients with
environment \citep{FLB:11}. Although the stellar populations
of ETGs are found to strongly correlate with velocity dispersion
\citep[see, e.g.][]{Bernardi:03}, or a similar
``local'' observable, environment-related variations are also found --
at fixed velocity dispersion -- using different samples and
methods \citep{Weinmann:06,Rogers:10,Peng:10,Smith:12,FLB:14}.  In
this regard, the SAMI survey provides a unique set to probe
environment-related trends, as, by construction, it comprises galaxies
in a field environment -- selected from the Galaxy and Mass Assembly survey, GAMA
\citep{GAMA:11} -- and a cluster
environment \citep[targeting eight low-redshift
clusters,][]{Owers:17}.

This paper focuses on the analysis of the radial gradients found in
the chemical composition and age of the stellar populations of
early-type galaxies. We will characterize the trends with respect to a
number of local\footnote{In this context, an observable is termed
'local' if it is defined for a given galaxy, as opposed to indicators
that relate to environment, i.e. extended over larger scales.} 
drivers of the star formation and chemical enrichment
processes, as well as consider variations with respect to environment
(cluster vs field/group). Section~\ref{Sec:Data} presents our working
sample of SAMI early-type galaxies, and Section~\ref{Sec:StPops}
describes the procedure followed to extract the stellar population
parameters. The derivation of radial gradients is outlined in
Section~\ref{Sec:Grads}. Our results are presented in
Section~\ref{Sec:Results}, and the interpretation of the trends, both
regarding the general sample and the separation of the trends with
respect to environment, are discussed in Section~\ref{Sec:Discussion}.
A concluding summary is given in Section~\ref{Sec:Summary}.

\begin{figure*}
 \raisebox{-0.5\height}{\includegraphics[width=100mm]{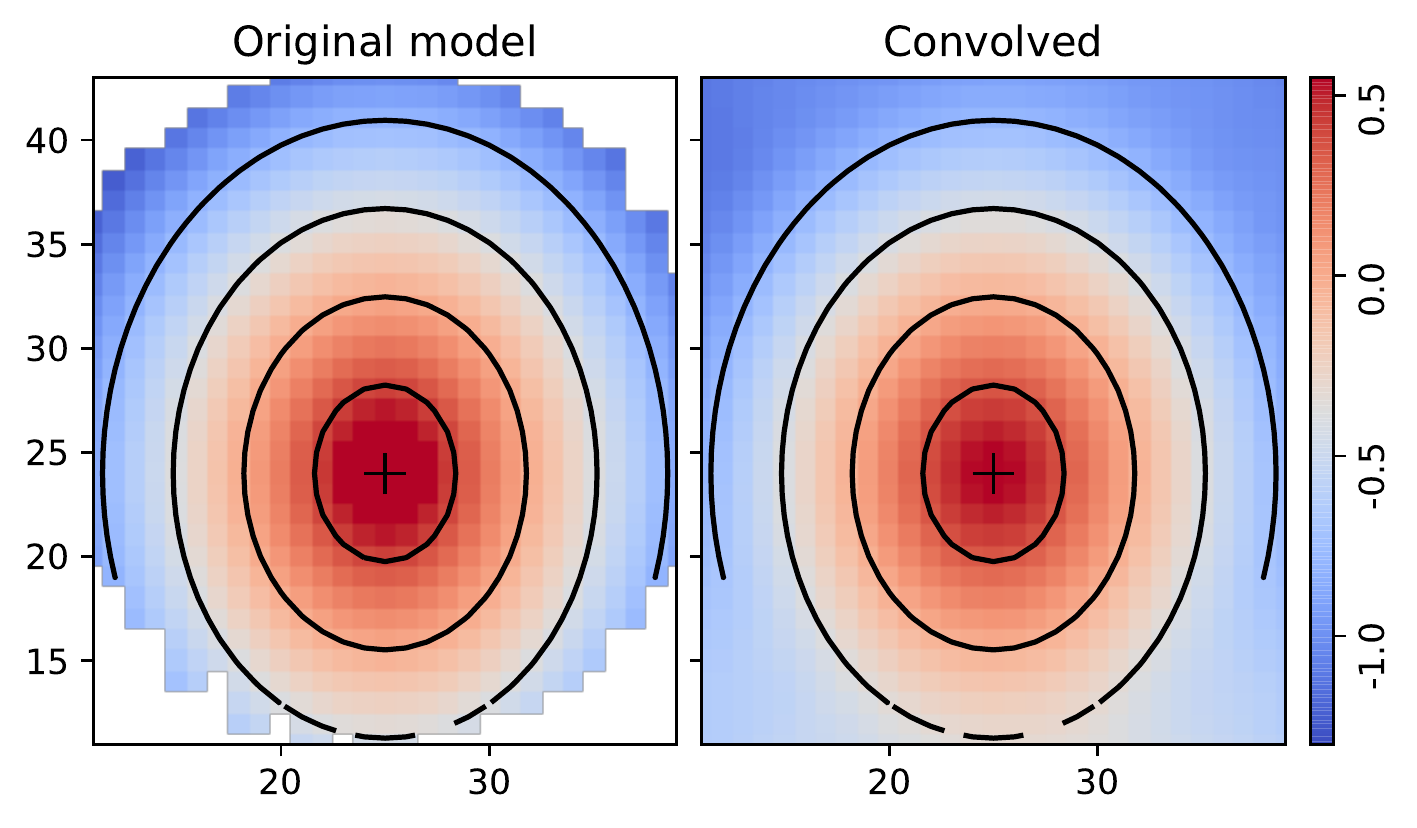}}
 \raisebox{-0.54\height}{\includegraphics[width=70mm]{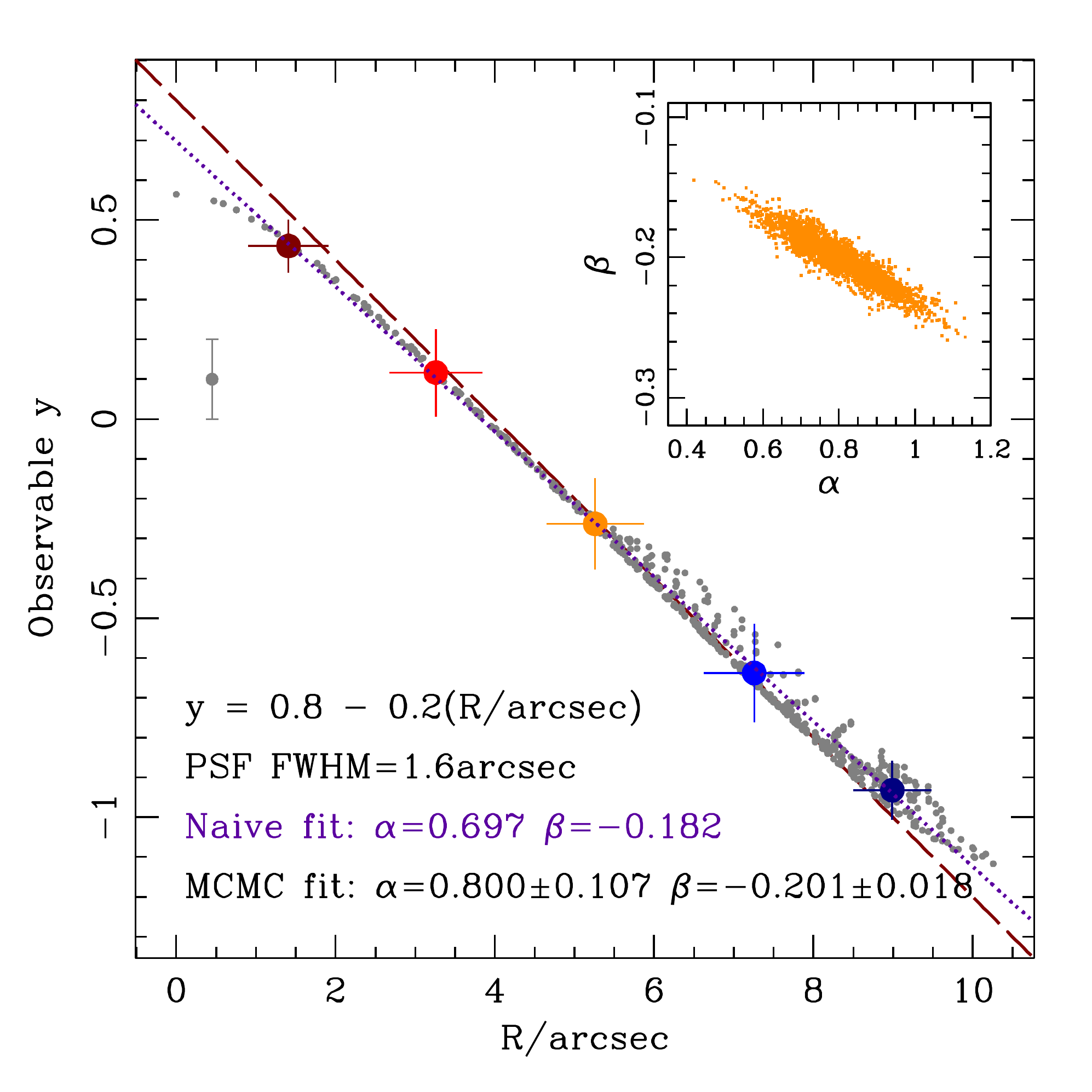}}
\caption{Illustration of the methodology applied to correct the contribution
of the Point Spread Function (PSF) in the derivation of radial gradients. The
left panels show a test case adopting a predefined model.
The contours follow the elliptical bins used
to derive the radial measurements. A model of the
parameter $y=0.8-0.2($R/arcsec$)$ is imposed, and the observed PSF
parameters are used to convolve the original model (leftmost panel)
into an observed one, shown in the adjacent panel. The panel on the
right shows the individual spaxel data (in grey), the radially binned
data (with error bars corresponding to the RMS scatter) and the
retrieved best fit for a linear model $y=\alpha + \beta($R/arcsec$)$
(dashed line) with slope $\beta=-0.201\pm 0.018$. A naive fit (not taking
into account the PSF) produces a shallower slope (dotted line,
$\beta=-0.182$).  The inset shows the distribution of parameters from the
MCMC sampler.
\label{fig:Conv}}
\end{figure*}

\section{Early-type galaxy sample}
\label{Sec:Data}

This work is based on data taken by the Sydney
Australian-Astronomical-Observatory Multi-object
Integral-Field-Spectrograph (SAMI). The development of hexabundle
technology \citep{hex1,hex2} enabled a generation of versatile Integral Field Units
(IFU) such as SAMI. Our starting sample comprises all galaxies classified with an
early-type morphology within the SAMI Galaxy Survey \citep{Croom:12}.
This survey consists of spectroscopic observations of $\sim$3,600
galaxies taken with the SAMI IFU at the 3.9\,m
Anglo-Australian Telescope. The SAMI Galaxy Survey is  described
in \citet{Bryant:15}, with additional details of the cluster sample
in \cite{Owers:17}. The full sample 
was visually inspected and morphologically classified as described
in \cite{Cortese:16}.  Galaxies were first divided into
spiral/non-spiral based on the presence of spiral arms or strong,
regular dust features, then further subdivided based on other
morphological features. We selected all galaxies classified as having
a visual morphological type of E, E/S0 or S0.
Our SAMI/ETG sample comprises 522 systems, split into 234 ETGs from
the GAMA survey (i.e. in a field/group environment) and 288 ETGs in
clusters. The sample covers a wide range of mass and size, and extends
over a redshift window z=[0.013,0.095], with a median value
z$_M$=0.053.  Fig.~\ref{fig:Sample1} shows the distribution with
respect to velocity dispersion ($\sigma$, {\sl left}) and stellar mass
(M$_s$, {\sl right}), both on a log scale.  Note the
difference in logarithmic range (0.5\,dex in $\sigma$ and 1.5\,dex in
M$_s$), and the scatter of the mean relationship, apparently larger in
velocity dispersion. Although no substantial bias is apparent between
GAMA and cluster galaxies, in Section~\ref{SSec:homog} we will
construct subsamples that remove any potential systematic caused by
the sample selection.

For each  galaxy, we  stack individual spaxel  spectra in  radial bins
following  the   elliptical  isophotes,   as  provided  in   the  SAMI
datacubes \citep{Nic:18}. A correction regarding the rotation velocity
is applied independently to each spaxel -- derived from the kinematics
analysis \citep{Jesse:17} -- to ``align'' all spectra to a common rest
frame,  before stacking  the  spaxel data  corresponding  to the  same
radial  bin. Typically,  the data  have between  3 and  5 radial  bins
available.  Fig.~\ref{fig:Sample2} shows  a comparison  of the  radial
extent of  the observations as a function of stellar mass
-- R$_{\rm  LAST}$ represents  the radial
extent of the  outermost radial bin.  The Point  Spread Function (PSF),
measured for each galaxy by the observation of a star in the same field, 
is quantified  by the Half-Width  at Half-Maximum (HWHM), to compare
it on equal terms with galaxy radii.
The median of the ratio R$_{\rm LAST}$/R$_e$ is 2.2. All radii
are quoted as circularized values: R$_e=\sqrt{a_eb_e}$, and the effective
radii are retrieved from S\'ersic fits to the surface
brightness profile (see \citealt{Kelvin:12} and \citealt{Owers:19} for details).

\section{Extracting population parameters}
\label{Sec:StPops}

In order to analyze the spectra, we rely on Simple Stellar Population (SSP) model predictions from 
\citet[][hereafter $\alpha$-MILES]{Vazdekis:15}. The $\alpha$-MILES SSPs are based
on the MILES stellar library, applying corrections from theoretical
models of stellar atmospheres to produce synthetic spectra of old- and
intermediate-age stellar populations at 2.51\,\AA\ (FWHM) spectral
resolution, with varying total metallicity (\zh ), IMF, and \afe\
abundance ratios. For the present work, we use models based on the
BaSTI (instead of the Padova) isochrones, as these are computed at
both [Mg/Fe]=0 (scaled-solar) and [Mg/Fe]=+0.4. We use SSPs with
metallicities\footnote{ Notice that we do not use $\alpha$-MILES models
with metallicity \zh$=+0.40$, as the corresponding predictions are
less safe (see V15 for details).  }
[Z/H]=\{$-0.96$, $-0.66$, $-0.35$, $-0.25$, $+0.06$, $+0.15$, $+0.26$\}, ages from 1 
to 14\,Gyr (with a 1\,Gyr sampling), and a Kroupa Universal IMF.
Line-strength predictions from the $\alpha$-MILES models are linearly
interpolated over a three-dimensional grid, with 100 equally-spaced
steps in both age and metallicity, and 175 steps in [Mg/Fe].  A linear
extrapolation is applied to extend the metallicity range up to
$+0.5$\,dex, and to probe the [Mg/Fe] range between $-0.15$ and
$+0.7$\,dex. However, we note that only in 5.9\% (1.9\%) of the cases 
it was necessary to invoke this extrapolation to obtain best-fit
values of the stellar populations in [Z/H] ([Mg/Fe]).

For each SAMI spectrum, we  estimated stellar population properties, namely
age, metallicity (\zh ), and \mgfe , by minimizing the standard $\chi^2$ statistic, namely:
\begin{equation}
\rm 
\chi^2 = \sum_i \left(\frac{O_i - M_i}{\sigma_i}\right)^2,
\label{eq:chi}
\end{equation}
where the index $\rm i$ runs over a selected set of spectral indices,
O$_i$(M$_i$) are the observed (model) line-strengths, and $\sigma_i$ are
the measurement errors of O$_i$.
We considered different sets of spectral indices, including two combinations of Balmer lines:
either \hbo\ only, or both \hbo\ and \hgF. For each set of Balmer lines we included all 
possible permutations of Fe indices, out of \fei , \feii , and \feiii . In each case,
we included \mgb , as it is required to constrain {\sl both} metallicity and  [Mg/Fe].
These indices are typically
measured as an equivalent width (EW), namely:
\begin{equation}
  {\rm EW}\equiv\int_{\lambda_1}^{\lambda_2}\left[
  1-\frac{\Phi(\lambda)}{\Phi_C(\lambda)}\right],
\end{equation}
where $\lambda_1$ and $\lambda_2$ define the central window of the
spectral feature, $\Phi(\lambda)$ denotes the spectrum under study and
$\Phi_C(\lambda)$ is the pseudo-continuum, given as a straight line
connecting a blue and a red sideband that straddle the central
feature. The uncertainty of the index is obtained by propagating the
corresponding uncertainty in the observed spectrum.  The central, blue
and red sidebands follow the standard definition, and are taken
from \citet{Trager98}, except for \hbo, defined in~\citet{CV09}. For
each spectrum, all model indices were computed after smoothing the
$\alpha$-MILES SSPs to match its effective broadening (instrumental
resolution and velocity dispersion).  The amount of broadening was
estimated with {\sc pPXF}~\citep{Cap:2004}, performing spectral
fitting in the rest-frame window 4,030--5,380\,\AA, that include all
the spectral features targeted here.  For each set of indices, we
determine the best-fitting stellar population properties by minimizing
Eq.~\ref{eq:chi} over the interpolated grid of $\alpha$-MILES SSP 
line strengths (with varying age, \zh , and \mgfe ). The parameter
uncertainties are derived following a Monte Carlo approach, producing
realizations of all indices when Gaussian noise, consistent with the uncertainty, 
is added to the line strengths.

\begin{figure*}
\begin{center}
\includegraphics[height=85mm]{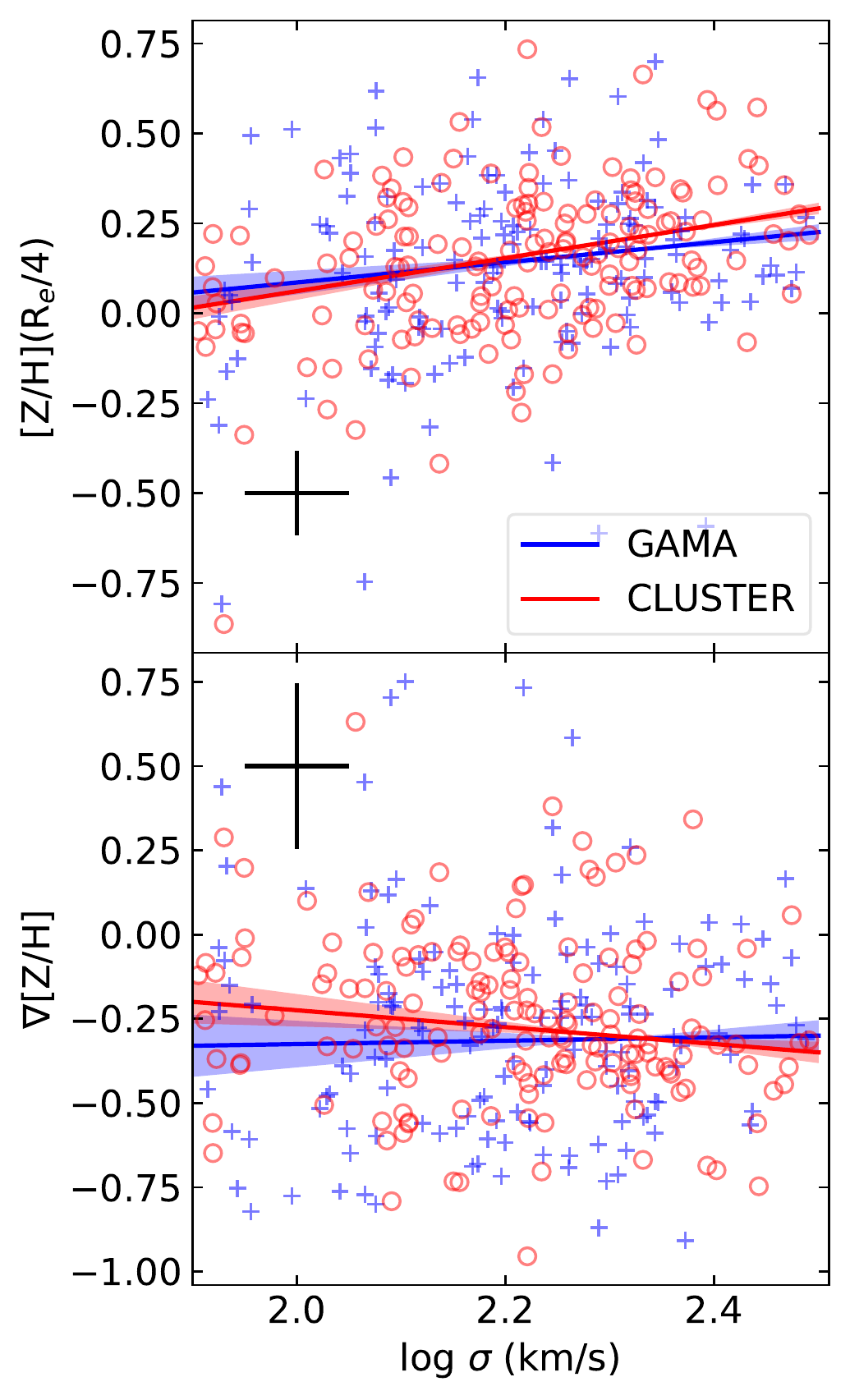}
\includegraphics[height=85mm]{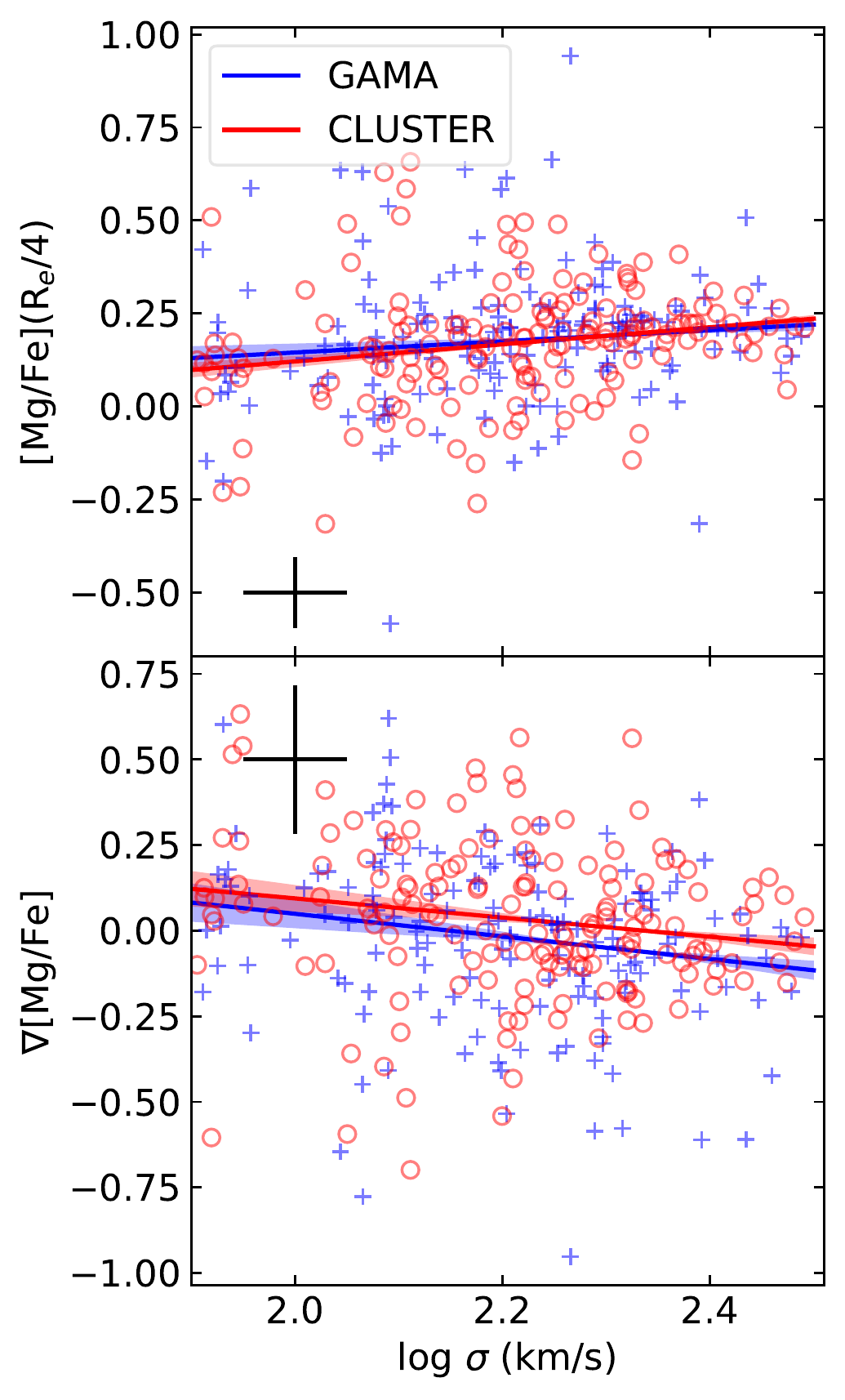}
\includegraphics[height=85mm]{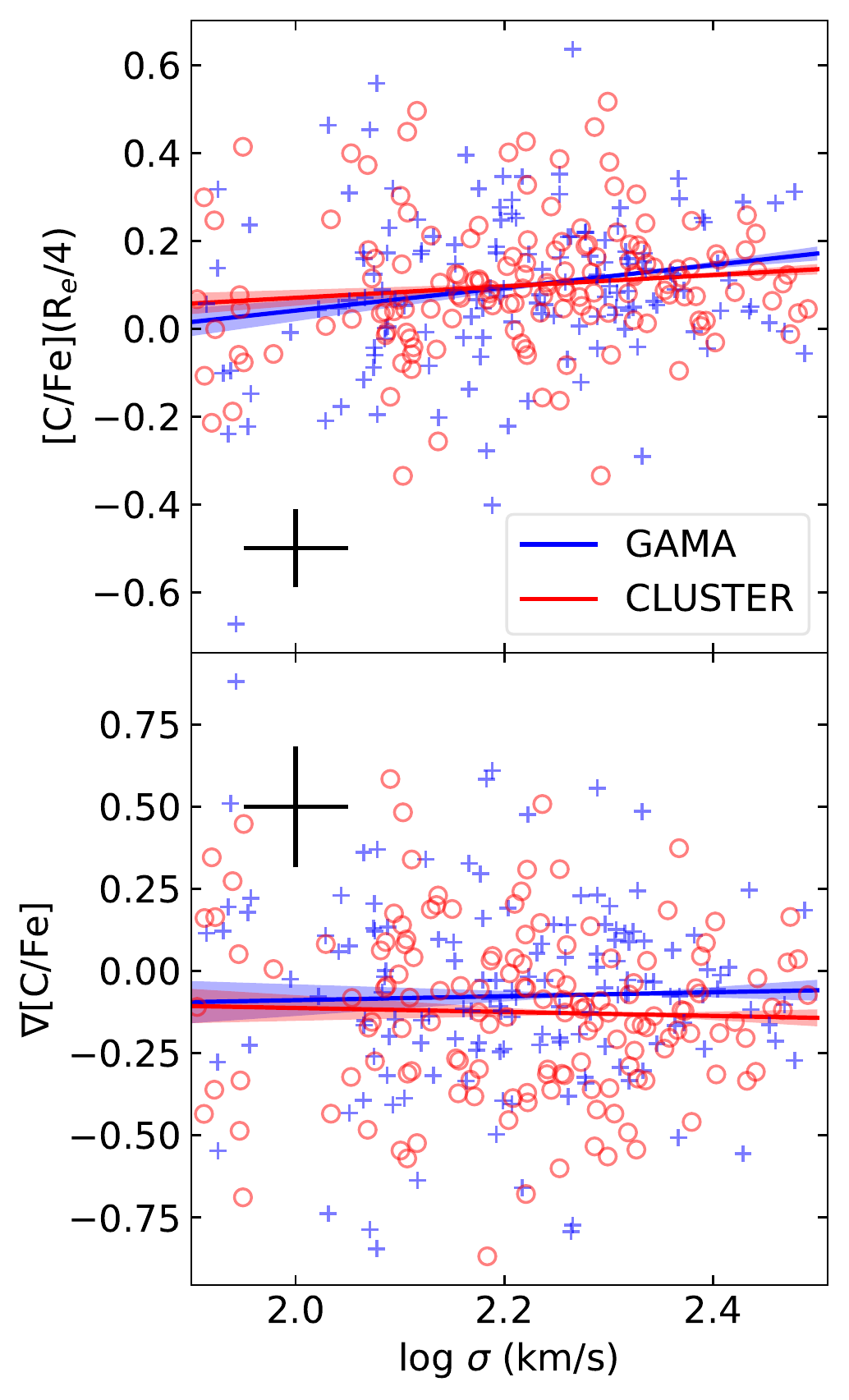}
\end{center}
\caption{Linear regression to the radial gradients (bottom panels) and
the intercept at $\sigma$=200\,\kms (top) of the (from left to right)
total metallicity, [Mg/Fe] and [C/Fe]. A typical error bar of an
individual measurement is shown in each panel, corresponding to the
median of the uncertainties in each case. The best fit results are
shown as solid blue (red) lines for the GAMA (cluster) subsample,
shown as blue crosses (red open circles).
\label{fig:Grads}}
\end{figure*}

In order to account for nebular contamination in the Balmer lines, we correct 
the \hbo\ and \hgF\ line-strengths with a similar procedure to that described 
in~\citet{LB:13}. We estimated the excess of flux in the line with respect 
to a combination of two SSPs, multiplied by a polynomial, giving the best fit 
in the \hbo\ (\hgF ) spectral region, 4,830--4,890\,\AA\ 
(4,310--4,370\,\AA), after excluding the absorption trough. 
The emission correction uncertainty was obtained by varying the degree, $N_p$, of the multiplicative 
polynomial in the fits (from $N_p=5$ to $11$), taking the standard deviation of 
the estimated emission corrections. The correction of \hbo\ turned 
out to be significant for $\sim 8 \, \%$ of the SAMI spectra,  with a median value of 
0.2\,\AA. Since the nebular contamination of \hgF\ is usually less significant than that 
on \hbo, we only needed to correct \hgF\ in $\sim$1\% of the spectra.

\begin{figure}
\begin{center}
\includegraphics[width=80mm]{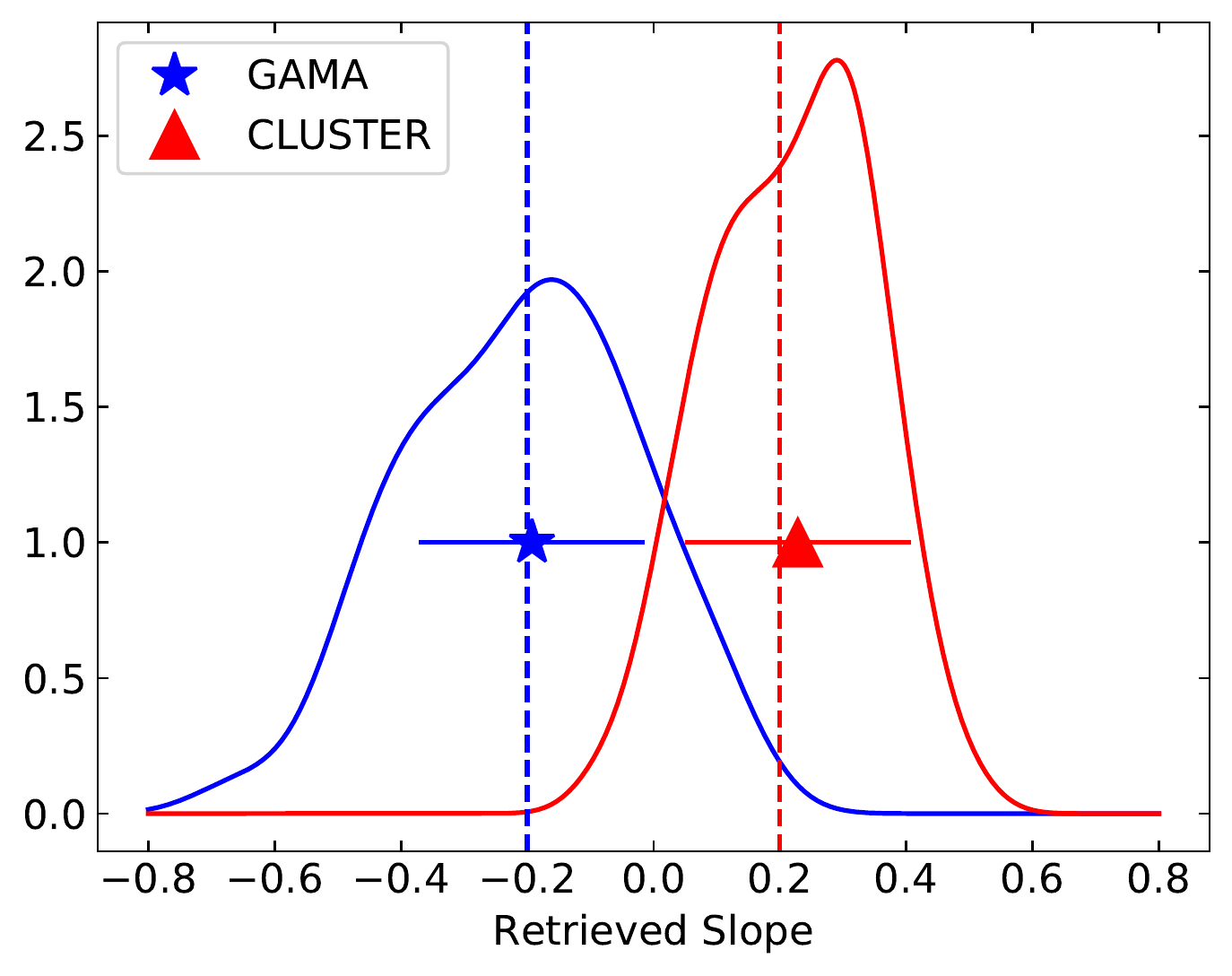}
\end{center}
\caption{Distribution of retrieved slopes in a synthetic set of data
with the same error distribution of the observed values of [Z/H],
enforcing a relation [Z/H]=$c\log\sigma_{200}$, with $c=-0.2$ for the
GAMA sample (blue) and $c=+0.2$ for the cluster sample (red). See text for
details.
\label{fig:Mocks}}
\end{figure}

\begin{table*}
\caption{Details of the homogeneous samples defined with respect to
the possible drivers. Cols.~1 and 2 identify the ``driver''
targeted in each case. 
The original sample of SAMI ETGs with
population gradient measurements comprises 211 GAMA galaxies and
245 cluster galaxies. The homogenised samples, by construction,
have equal number of galaxies from GAMA and from clusters, labelled $N$ in
col.~3. Col.~4 is the Kolmogorov-Smirnov statistic $D_{KS}$ for the
original sample, whereas col.~5 gives the equivalent when comparing
the homogenised subsamples. To test the significance, col.~6 gives
the mean and standard deviation of the distribution of $D_{KS}$ for
1,000 random reshufflings of the data.
\label{tab:homog}}
\begin{center}
\begin{tabular}{rllcccc}
\hline
\multicolumn{2}{|c|}{Driver} & Definition & N & $D_{KS}^{\rm orig}$ & $D_{KS}^{\rm homog}$ & $D_{KS}^{\rm random}$\\
\multicolumn{2}{|c|}{(1)} & (2) & (3) & (4) & (5) & (6)\\
\hline
  I& Velocity dispersion ($\sigma$)  & $\delta_1=\log(\sigma/200\,{\rm km\,s}^{-1})$ & $178$ & $0.125$ & $0.067$ & $0.064\pm 0.020$\\
 II& Stellar mass ($M_s$)  & $\delta_2=\log(M_s/10^{11}M_\odot)$ & $157$ & $0.278$ & $0.072$ & $0.069\pm 0.022$\\
III& Dynamical Mass ($M_d$) & $\delta_3=\log(M_d/10^{11}M_\odot)$ & $174$ & $0.208$ & $0.081$  & $0.067\pm 0.021$\\
 IV& Surface stellar mass density ($\Sigma_s$) & $\delta_4=1+\delta_2-2\log($R$_e/2\,{\rm kpc})-\log 2\pi$ & $200$ & $0.063$ & $0.056$ & $0.062\pm 0.020$\\
  V& Stellar potential ($\Phi_s$) & $\delta_5=\delta_2-\log($R$_e/2\,{\rm kpc})$ & $172$ & $0.230$ & $0.057$ & $0.067\pm 0.020$\\
 VI& Virial test ($\sigma^2/R$) & $\delta_6=\delta_1-\log($R$_e/2\,{\rm kpc})$ & $177$ & $0.181$ & $0.057$ & $0.065\pm 0.021$\\
\hline
\end{tabular}
\end{center}
\end{table*}

Some of the SAMI spectra were significantly contaminated by sky line 
subtraction residuals, especially in the outermost radially-binned
spectra (corresponding to lower surface brightness levels). Different
lines were affected, depending on the redshift of the galaxy.  In
order to tackle this issue, we flagged out contaminated features --
for each of the radially-binned spectra -- by comparing the observed
spectra with the best-fitting ones, obtained with {\sc pPXF} (see
above).  For each galaxy, we only considered the results obtained from
sets of spectral indices not affected by sky residuals, averaging out
the corresponding best-fitting parameters (age, \zh , and \mgfe).  For
galaxies for which no spectral features were flagged out, we verified
that different sets of spectral indices produced, on average,
consistent results, justifying our approach.

We also estimated the \cfe\ abundance ratio, based
on the \cfs\ spectral index~\citep{Trager98}, mostly sensitive to the
carbon abundance. Given the best-fitting age, \zh , and \mgfe\ (see
above), we use the $\alpha$-MILES models to derive the corresponding
value, \cfs$_{\rm M}$.  We then obtained \cfe\ from the following
equation:
\begin{equation}
\rm 
[C/Fe] = \frac{ ( C4668 - C4668_{\rm M} ) }{ C4668_{\rm M} \cdot S_{C}},
\label{eq:cfe}
\end{equation}
where $\rm S_C$ is the relative sensitivity of the \cfs\ index to \cfe
, i.e.  $\rm S_C = \delta( C4668) / C4668 / [C/Fe]$.
We computed $\rm S_C$ from \citet{CvD12} stellar population models, for an age of 13.5~Gyr,
solar metallicity, and a Chabrier IMF. Notice that in this approach, we
assume a constant $\rm S_C$, i.e. independent of age, \zh ,
and \mgfe .  We tested the applicability of this approach by use of the 
\citet{TMJ11} stellar population
models\footnote{These models provide predictions for \cfs\ at varying
age, \zh, \mgfe, and $ \rm  C / \alpha$. We verified that, for a wide range
of these parameters,  the value of $\rm S_{\rm C}$ is reasonably 
constant (within $\sim$20\,\%).}.

\begin{figure*}
\begin{center}
\includegraphics[width=150mm]{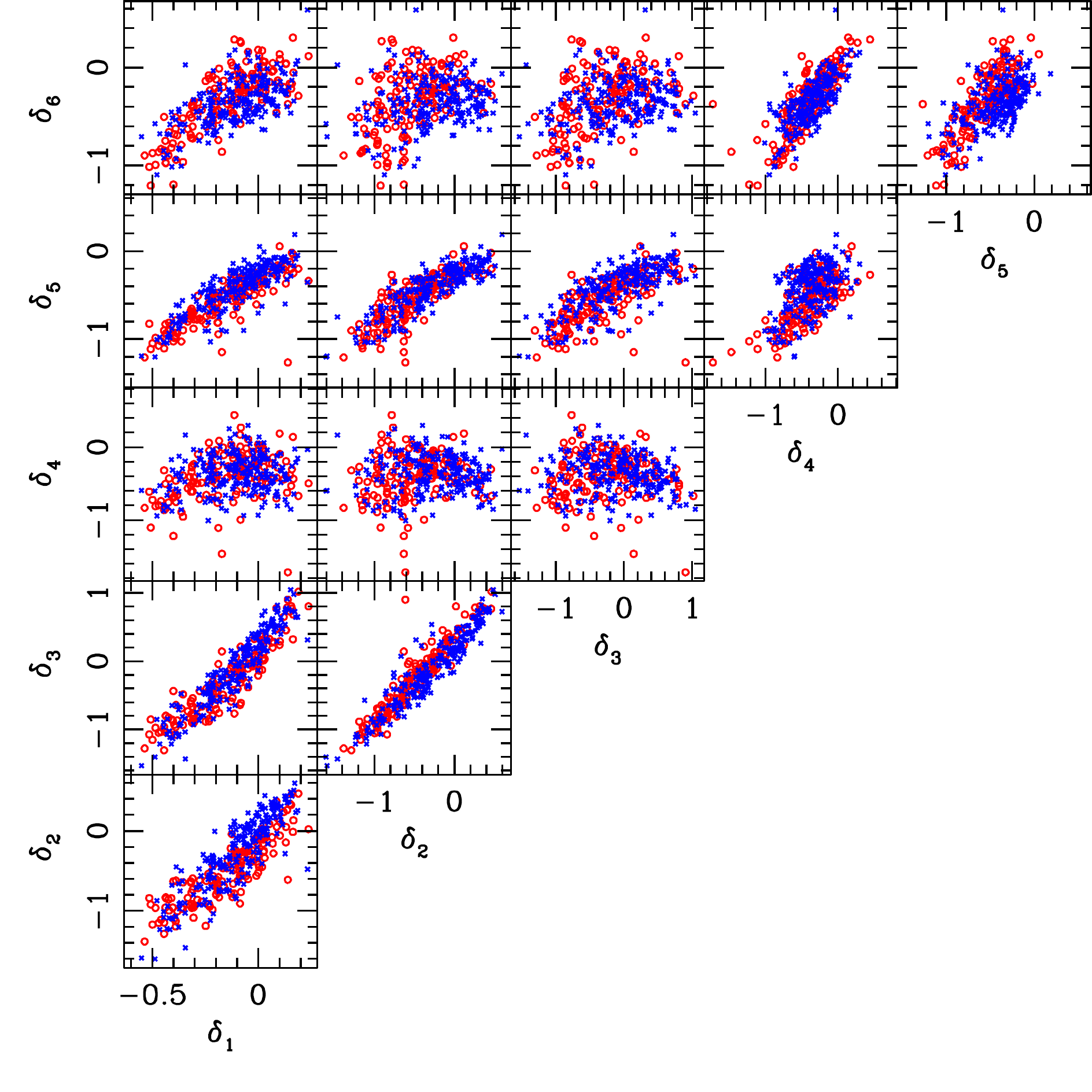}
\end{center}
\caption{Correlations among the drivers explored in this paper. The six drivers
$\{\delta_1,\cdots,\delta_6\}$ are defined in Table~\ref{tab:homog}. The notation of
the symbols is the same as in Fig.~\ref{fig:Sample1}
\label{fig:drivers}}
\end{figure*}

\section{Derivation of gradients}
\label{Sec:Grads}

Fig.~\ref{fig:Sample2} illustrates the spatial resolution limit of our sample,
where the effective radius, the last radial bin available from the IFU data,
and the (HWHM) extent of the PSF, are compared. In most cases, the PSF width is
significant, although it does not represent a major drawback: the median value
of R$_{\rm LAST}$/PSF$_{\rm HWHM}$ is 6.2, whereas the median of R$_e$/PSF$_{\rm HWHM}$
is 2.5. Nevertheless, we need to account for the spatial resolution limit in order
to extract robust gradients, with meaningful error bars. We
apply a forward modelling methodology as follows.
Let us assume that $\{y_i\}$ represents a set of population
parameters, derived from the analysis presented in \S\ref{Sec:StPops}.
These observations correspond to the individual radial bins $\{R_i\}$.
We fit the results to a model given by a linear function $y=\alpha+\beta R$, and
populate the individual spaxels of the observation with this (exact) model.
The model is convolved with the PSF of that 
observation, defined by a Moffat profile, with parameters taken from the
FITS header of each observation (measured from the observation of a nearby
star during the same exposures). The convolved model
is then mapped on to the layout of the radial bins, in order to
create a set of binned parameters $\{y^M_i\}$, that are
compared with the original observations, following a standard likelihood based on a
$\chi^2$ statistic:
\begin{equation}
\ln{\cal L} = \ln{\cal N} - \frac{\chi^2}{2},
\end{equation}
where we follow the standard definition:
\begin{equation}
\chi^2 = \sum_i\left(\frac{y_i-y^M_i}{\sigma_i}\right)^2,
\end{equation}
and $\sigma_i$ is the uncertainty corresponding to the
derivation of the population parameter $y_i$.
This process is implemented with an off-the-shelf MCMC sampler
\citep[{\sc emcee},][]{emcee}, to retrieve the best fit values of the
slope ($\beta$) and intercept ($\alpha$), along with their
uncertainties. This method is illustrated in Fig.~\ref{fig:Conv}
for one of the galaxies in our sample (ID 23623), whose data provide five
radial apertures, and the effective radius extends over $\sim$10 
times the HWHM of the PSF. The panels on
the left show a test model ($y=0.8 - 0.2R$, where $R$ is the circularised
radius of each annulus in arcsec) both before and after convolution. The
radial bins are overlaid, for reference. The panel on the right shows the derivation
of this test case. The grey dots
are the individual (i.e. spaxel) observations, and the coloured dots
are the radially binned measurements, including error bars that represent the
scatter within each annulus. The dotted line traces a naive least
squares fit not taking into account the effect of the PSF. It gives 
a slightly shallower gradient ($\beta=-0.182$), as expected, since the PSF tends to
wash out any potential gradient. The method presented here gives
an unbiased gradient ($\beta=-0.201\pm 0.018$), and the inset shows the
potential covariance between slope and intercept. We note that this
mock observation is created with comparable uncertainties to the
actual observations.

\section{Results}
\label{Sec:Results}

Fig.~\ref{fig:Grads} shows the distribution of gradients in chemical
composition with respect to velocity dispersion, from left to right, 
total metallicity ([Z/H]), [Mg/Fe], and [C/Fe].  GAMA ETGs are represented by
blue `+' symbols, and cluster galaxies appear as red `o' symbols,
respectively. The top panels show the value of the best linear fit at
one quarter of the effective radius (R$_e$/4), and the bottom panels
give the radial gradient, measured with respect to $\log$\,R, e.g.
$\nabla$[Z/H]$\equiv$d[Z/H]/d$\log$\,R. A typical error bar for the
individual measurements is shown in each panel.  The best fit appears
as a blue (red) line for GAMA (cluster) galaxies, including a shaded
region that spans the 1\,$\sigma$ uncertainty in the slope
of the fits (the uncertainty in the intercept of these fits,
defined as the value of the fit at $\sigma$=200\,\kms, is
substantially smaller than that of the slope).  The top panels of
the figure show the characteristic increase of metallicity and
abundance ratios with respect to velocity dispersion
\citep[e.g.][]{Trager:00}.

The accuracy of the method is tested on simulated data, by performing a set
of 100 Monte Carlo realizations that enforce a correlation
$[Z/H]=c\log\sigma_{200}$, with $c_G=-0.2$ for the GAMA set and $c_C=+0.2$ for
the cluster set. Each realization features the same number of galaxies as the
original set, with slopes retrieved from a Gaussian probability distribution
with mean $c_G$ or $c_C$, and standard deviation corresponding to the 
uncertainties of the observed data. In this way, we make sure the 
 simulated data has the same 
 distribution of uncertainties
as the original sample. Figure~\ref{fig:Mocks} shows the distribution
of values for the whole set of 100 realizations, with the mean and
standard deviation given by the symbols and error bars. The input slopes
($c_G$ and $c_C$) are 
represented by the vertical dashed lines, and the distribution of
measured gradients shows that our method is
fully consistent. Moreover, we compared the {\sl individual} estimates of the
slope uncertainty, produced for each realization, with the width of the
distribution of slope measurements, and obtained fully consistent results:
the GAMA set gave a median slope of $-0.183\pm 0.178$  whereas
the median of the individual uncertainties was $0.175$; the cluster
sample gave a median slope of $+0.218\pm 0.178$ and a median uncertainty
of $0.155$ (all results quoted at the 1\,$\sigma$ confidence level).

\begin{figure*}
\begin{center}
\includegraphics[height=85mm]{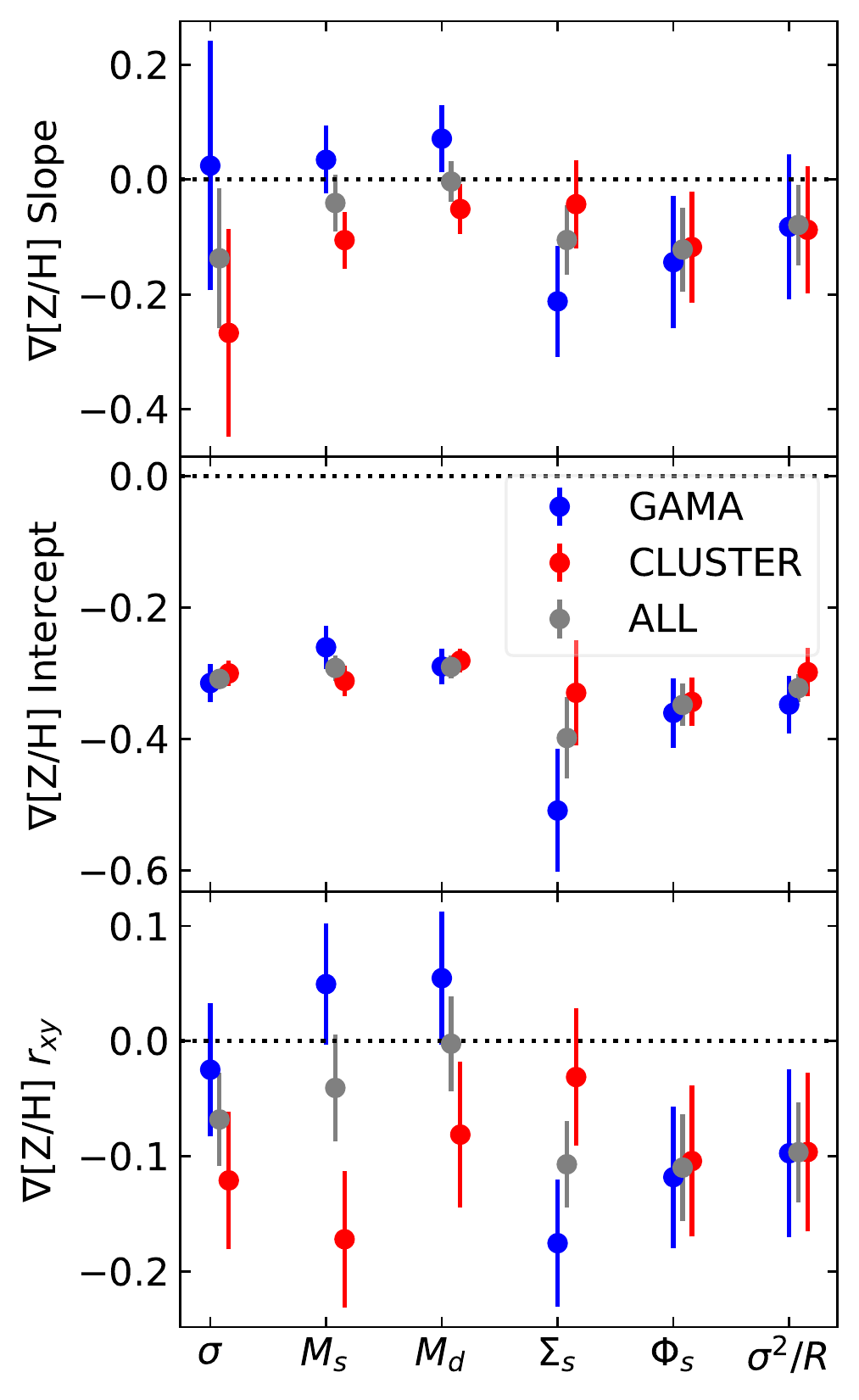}
\includegraphics[height=85mm]{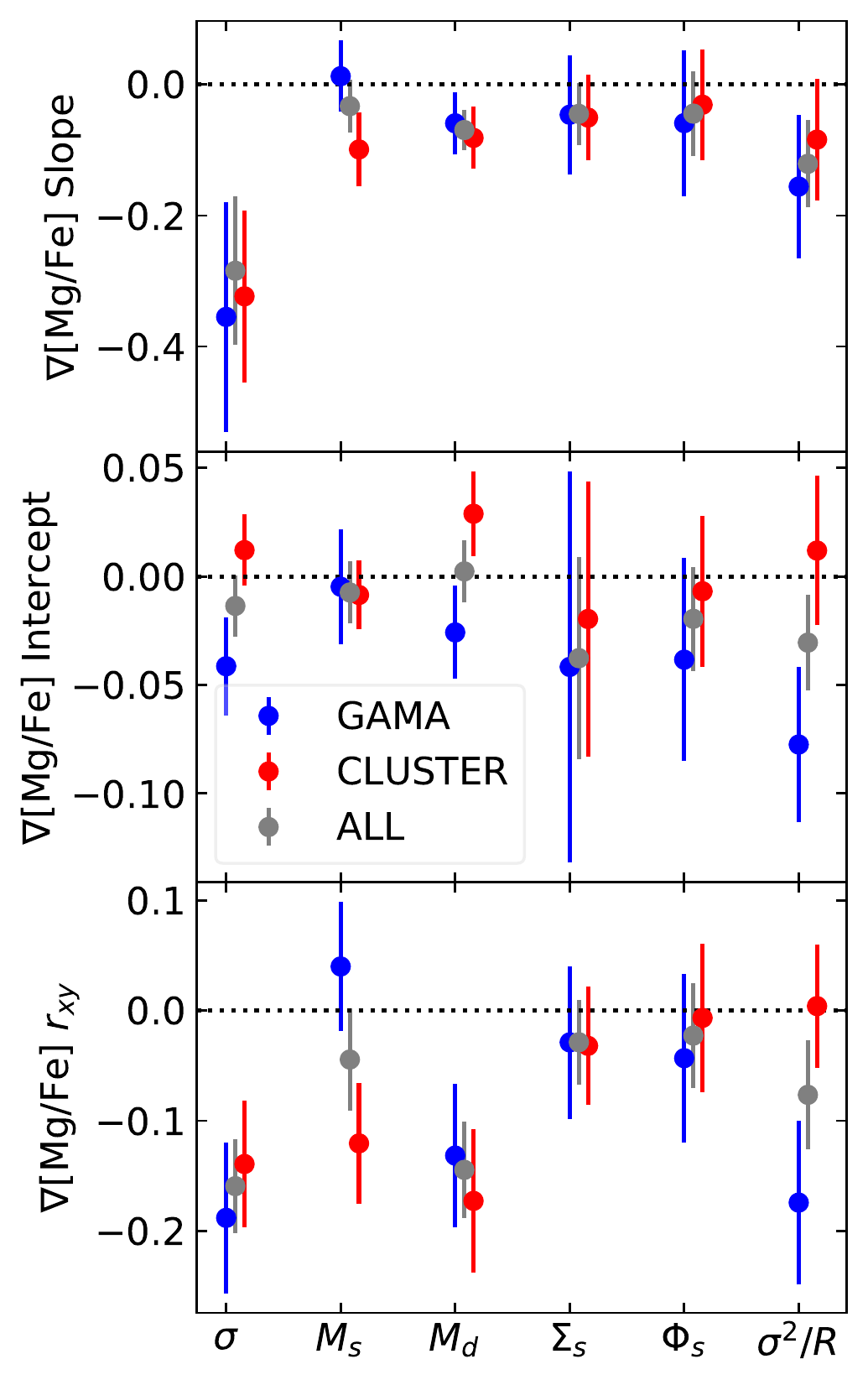}
\includegraphics[height=85mm]{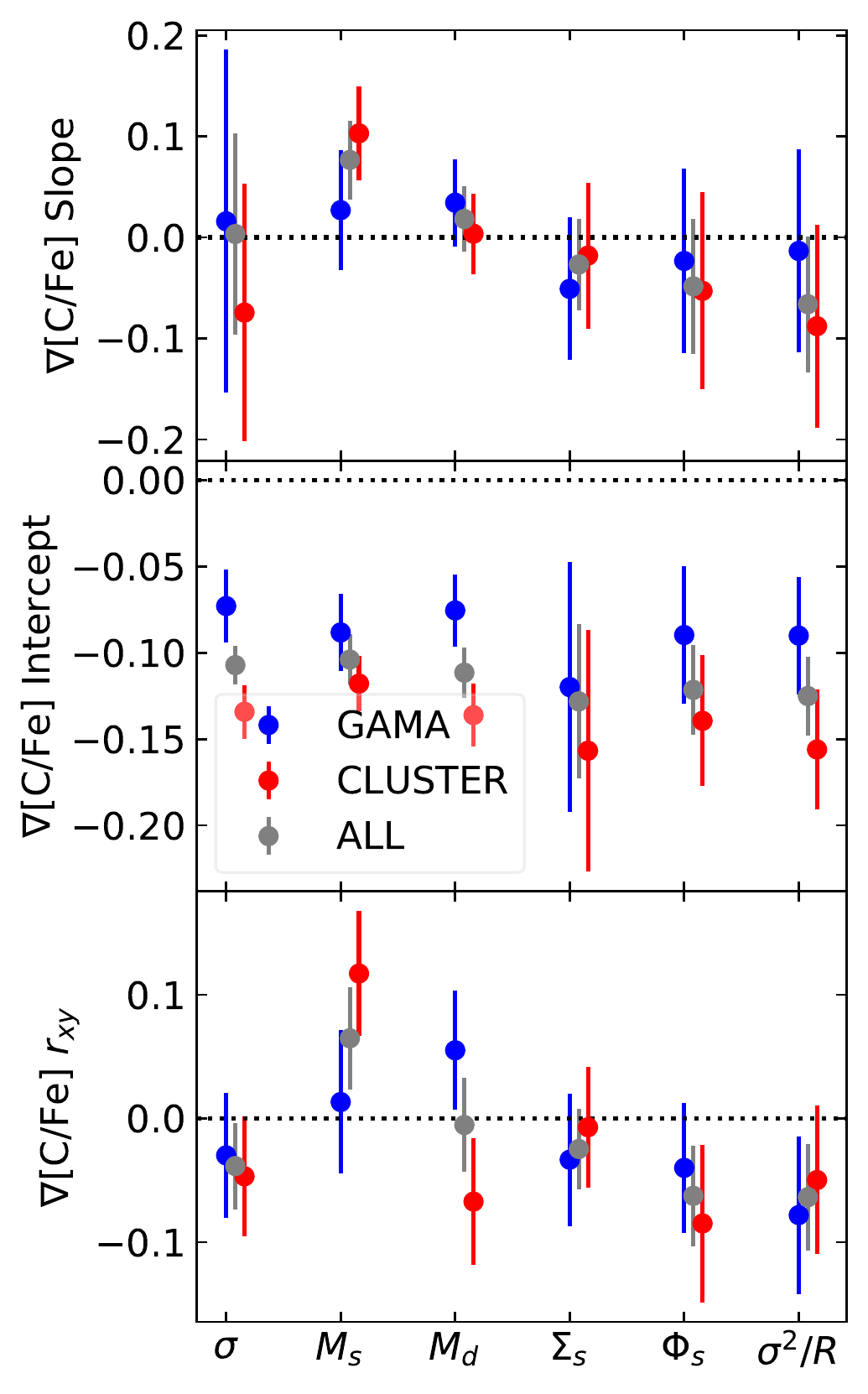}
\end{center}
\caption{Exploring the potential drivers of population gradients for the
trend of the radial gradient ($\nabla X\equiv\Delta X/\Delta\log R$).
\label{fig:Drivers_n}}
\end{figure*}

\subsection{Possible drivers of the observed trends}
\label{SSec:drivers}

The observed radial gradients of the stellar populations encode the
underlying formation process, including in-situ and ex-situ growth.
Such a complex and entangled mixture of possible contributors can only be
explored in a meaningful way if we scrutinise a reduced number of indicators
we define here as ``drivers'' of the observed trends. 

The first two columns of Table~\ref{tab:homog} define the set of six
drivers adopted in this work. Our criterion for this choice 
is to identify observables that are relatively easy to measure from
the photometric and spectroscopic properties of the
sample. Motivated by the recent work on
SAMI data and elsewhere \citep{Barone:18,dEug:18}, we choose the
velocity dispersion (measured as averaged within an effective radius);
the stellar mass; the dynamical mass; the
average surface mass density (defined as
$\log\Sigma_M=\log M_s-2\log$R$_e$); and the average gravitational potential when considering only
the stellar mass (again simply defined as $\log\Phi=\log M_s-\log R$,
i.e.  disregarding dark matter and assuming a homologous distribution
of matter). Note that the full gravitational potential -- i.e. involving
the total mass -- is $\Phi\propto\sigma^2$, so that our first driver
(velocity dispersion) can be considered a proxy of the total
potential\footnote{Note, however, that the quoted slopes corresponding
to $\Phi$ differ from those measured with respect to $\sigma$ by a factor 1/2.}
Fig.~\ref{fig:drivers} compares the distribution of these
drivers among one another, following the same notation as in
Fig.~\ref{fig:Sample1}. Note the strong correlation among several
pairs of drivers, such as $\delta_2$ (stellar mass) and $\delta_3$ (dynamical
mass), noting that these two masses are derived from independent observables.
In addition to the five drivers described above, we add a new one, $\delta_6$,
defined as $\sigma^2/$R$_e$. If we assume fully virialised systems,
this driver maps the total surface mass density, and, as expected,  $\delta_6$
correlates well with the surface stellar mass density ($\delta_4$).

Table~\ref{tab:fits} (in the Appendix) quantifies the slope, intercept
and linear correlation coefficient of all the fits to the data,
including the full set of ETGs, as well as the subsamples segregated
with respect to environment (see Section \ref{SSec:homog} for further details
about the homogenisation process applied, to minimise a bias in this
regard).  The error bars are quoted at the 1\,$\sigma$ level. We note
that there are two types of variations studied here: the radial
gradient of a given observable for an individual galaxy (i.e. each of
the data points in the bottom panels of Fig.~\ref{fig:Grads}) and
the correlation of the best fit values with respect to a ``driver'', such
as the velocity dispersion (i.e. the slopes of the lines in
Fig.~\ref{fig:Grads}). To avoid confusion, we refer to the former as
``gradient'', and the latter as ``slope''.

\subsection{Homogenising the samples}
\label{SSec:homog}

In addition to the general analysis concerning trends of population
gradients with respect to a number of possible physical drivers, as
presented above, we also look for differences regarding environment,
by comparing the GAMA sample -- that represents a general field sample
-- and the cluster sample specifically targeted in the SAMI survey.
However, differences in the distribution of the parameter under study,
say velocity dispersion, between the cluster and the field sample could
create a spurious difference that would be wrongly identified as an
environment-related effect. To avoid this issue, we need to produce 
``homogeneous'' subsets of GAMA and cluster galaxies that 
enforce the same distribution of the parameter being considered.
This approach improves over a mass-function
weighted analysis -- implemented in, e.g., \citet{Barone:18} -- by specifically
constructing samples which, as far as the chosen driver is concerned,
are undistinguishable. Firstly we define the target
distribution as the one corresponding to the total sample
with respect to the chosen driver, say velocity dispersion. Then, for
either subsample (cluster or GAMA), we randomly select galaxies within
a relatively narrow interval of this driver\footnote{The full range
of the parameter under study, spanned by the total sample of ETGs, is
binned into 16 intervals, within which the ratios of galaxy numbers
are enforced to be the same in the cluster and GAMA subsamples.},
enforcing this subsample to have the same distribution as the target
one. Of course, the drawback of this method is that a number of
galaxies have to be removed from the analysis to make sure the ``shape
of the histogram'' is the same in both subsets. However, no
significant variations are found in different realizations.

\begin{figure*}
\begin{center}
\includegraphics[height=85mm]{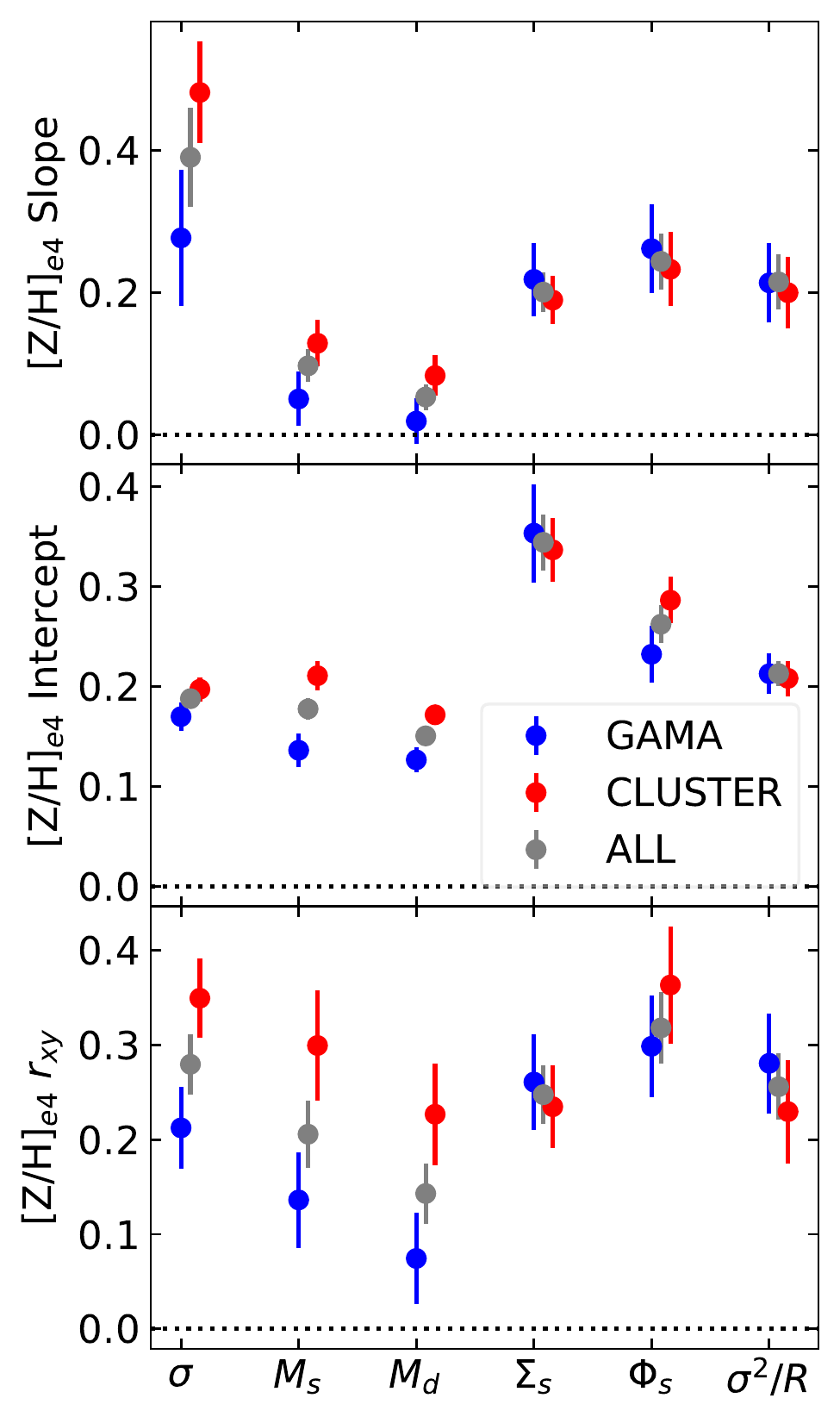}
\includegraphics[height=85mm]{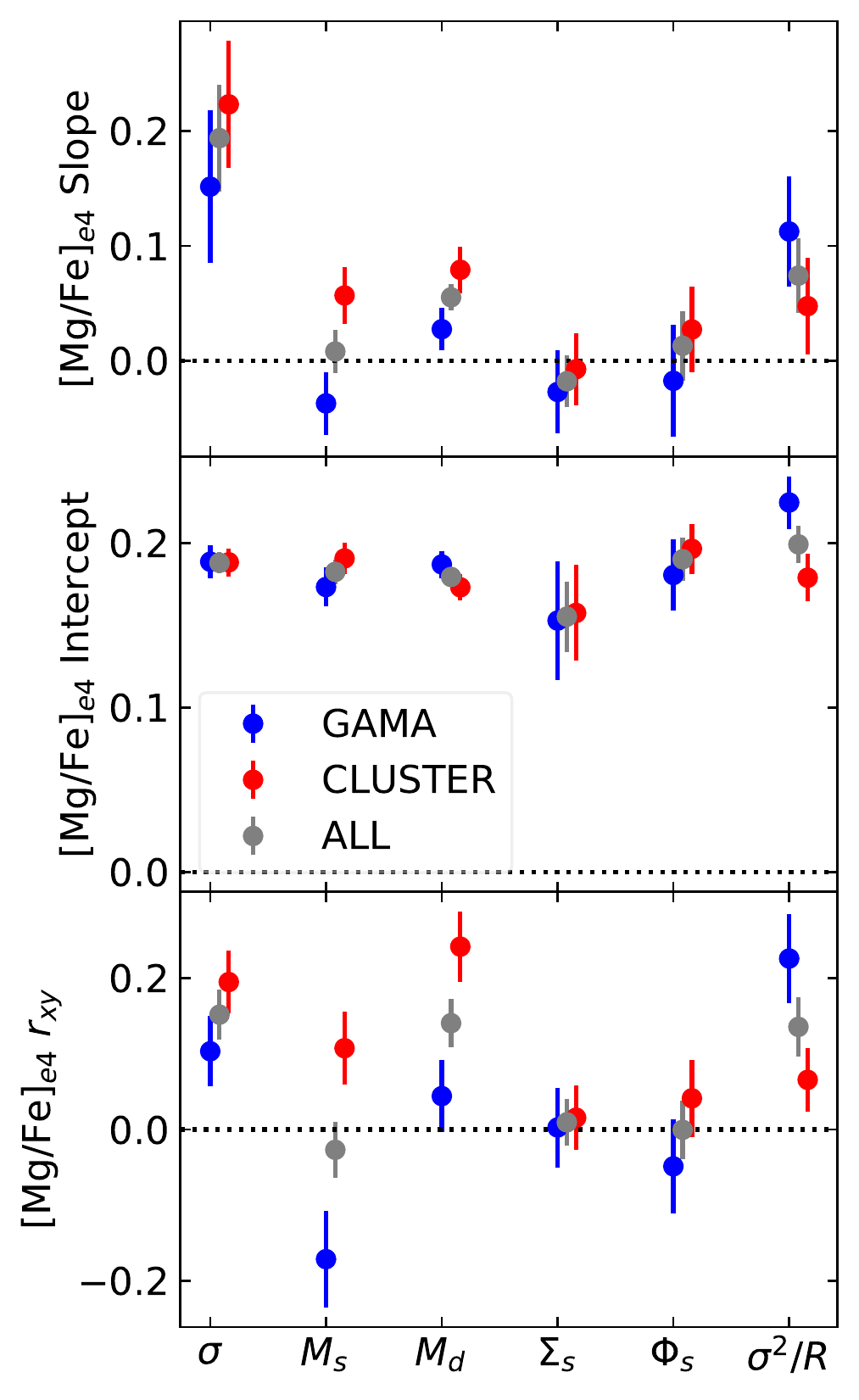}
\includegraphics[height=85mm]{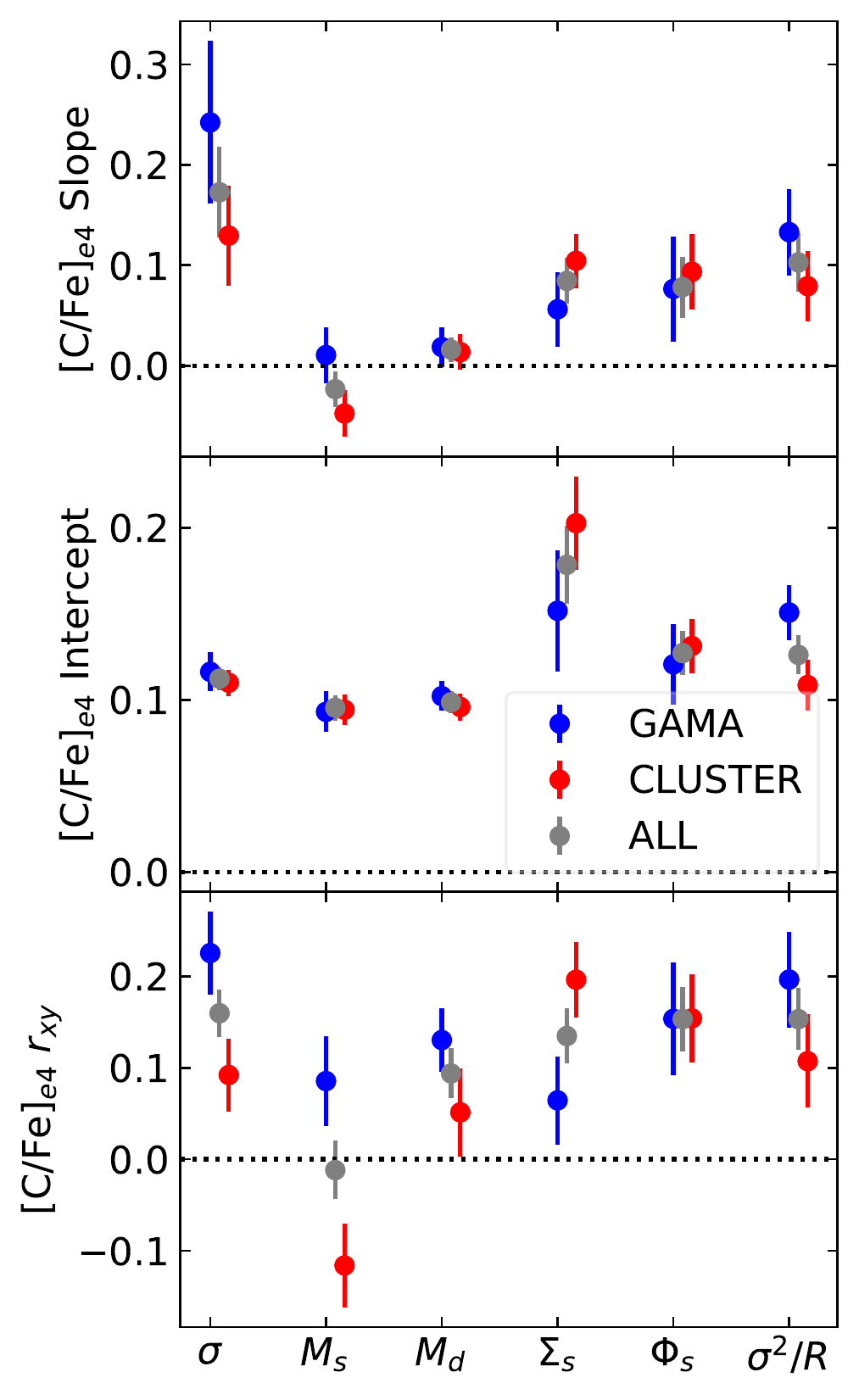}
\end{center}
\caption{Exploring a number of possible drivers of population gradients for the
trend of the ``central'' values (corresponding to the linear fit estimated
at R$_e$/4).
\label{fig:Drivers_e}}
\end{figure*}

Table~\ref{tab:homog} shows the statistical differences measured 
between the original sample and the homogenised one.
For each driver, we give the final number of galaxies in each subsample
(being equal, by construction), the KS statistic ($D_{KS}$) for the original and
homogeneous samples, and the KS statistic corresponding to a fully
random set: this one is obtained by randomly
reshuffling the targeted samples 1,000 times, producing a distribution
of D$_{KS}$ from which the mean and standard deviation are quoted.

The results of the slopes and intercepts for the whole set of six 
drivers are show on Tables~\ref{tab:fits} and \ref{tab:fits2}.
A linear regression is applied both to the radial gradients, and the
central values of the parameters.
The models for stellar parameter $\pi$ are thus $\nabla\pi=a\delta + b$
(for the gradient) and $[\pi]_{\rm e4}=a\delta + b$ (for the central value),
where $\delta$ is one of the six drivers defined above, and $\pi$
corresponds to either total metallicity ([Z/H]), [Mg/Fe], [C/Fe] or (log) stellar
age.  $r_{xy}$ is the linear correlation
coefficient. The error bars, quoted at the 1\,$\sigma$ level, take
into account the individual uncertainties of the measurements. The
intercepts are given by the $b$ coefficients, and correspond to 
the population parameter at a reference value
of the driver, as shown in col.~2 of Table~\ref{tab:homog}. These
reference values adopt a fiducial galaxy with 
velocity dispersion $\sigma$=200\,\kms, stellar (or
dynamical) mass M$_s=1\times 10^{11}$M$_\odot$, and effective radius
R$_e$=2\,kpc.

\section{Discussion}
\label{Sec:Discussion}

We split the discussion into a general analysis of the trends
with respect to the drivers defined above (Table~\ref{tab:homog}),
followed by a comparison of the results with respect to environment, i.e.
contrasting the GAMA and cluster subsamples.

\begin{figure*}
\begin{center}
\includegraphics[height=85mm]{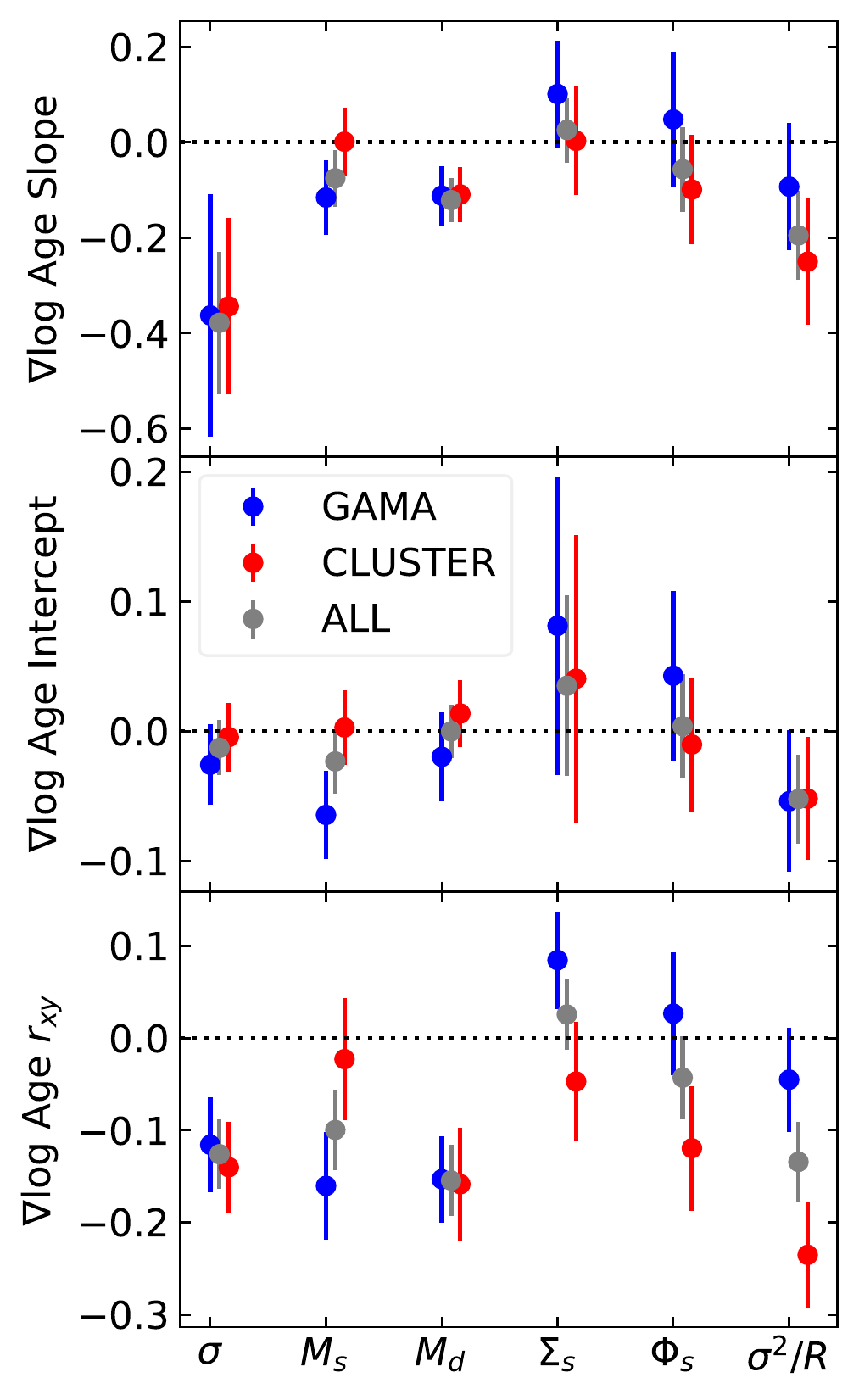}
\includegraphics[height=85mm]{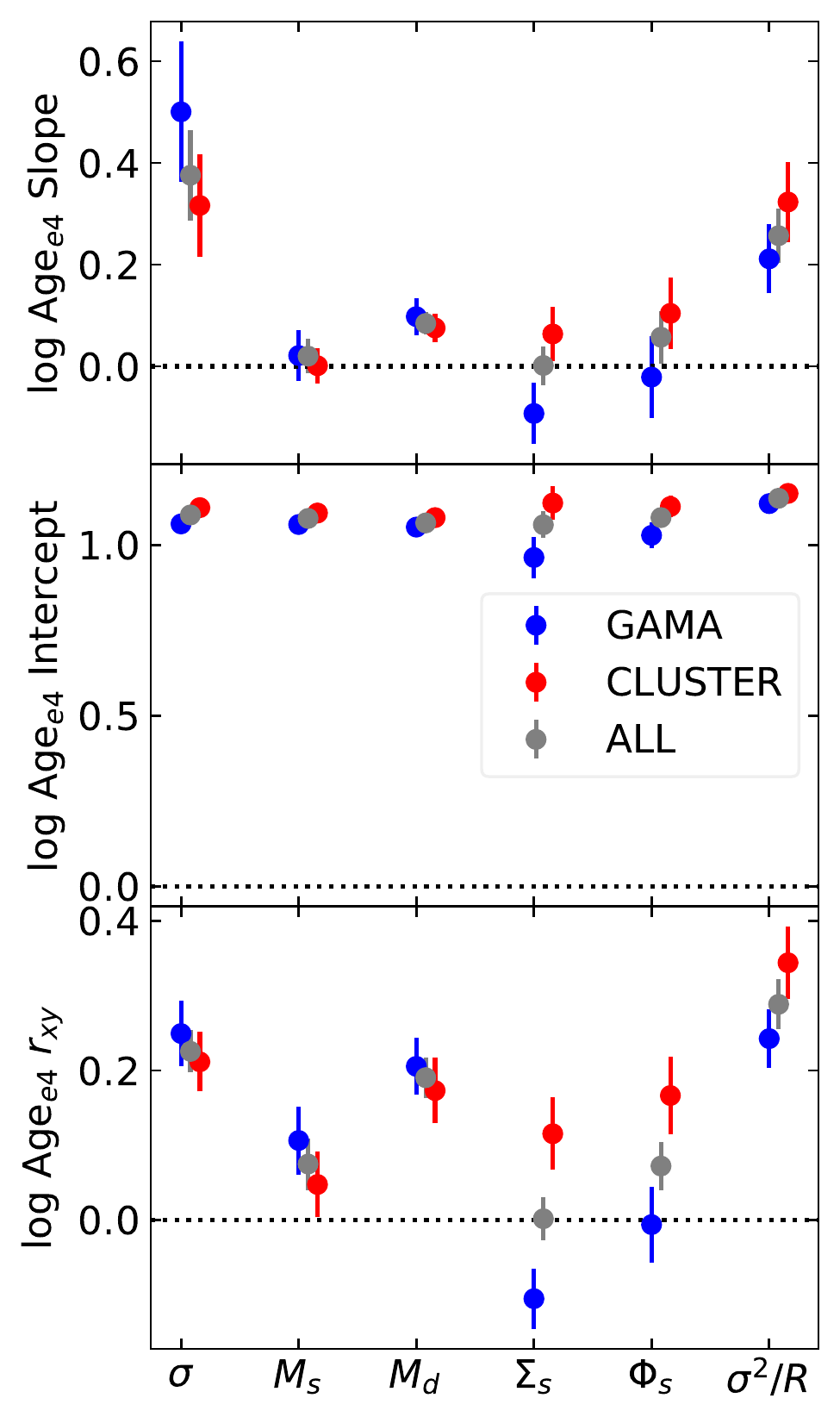}
\end{center}
\caption{Equivalent version of Figs.~\ref{fig:Drivers_n} (left, radial gradient trends) and
\ref{fig:Drivers_e} (right, central value trends),
regarding (log) stellar ages.
\label{fig:Drivers_lAge}}
\end{figure*}

The results obtained with respect to the different drivers
are shown in Figs.~\ref{fig:Drivers_n}, \ref{fig:Drivers_e} and
\ref{fig:Drivers_lAge}, and quantified in Tables~\ref{tab:fits}
and \ref{tab:fits2}. The information for each observable 
is presented in three vertical panels: the top one gives the slope
of the (linear) trend corresponding to the chosen observable with respect
to the driver labelled in the horizontal axis; the middle
panel is the intercept of this linear trend, estimated at the
reference value of the driver (as shown on col.~2 of Table~\ref{tab:homog}),
and the bottom panel is the linear correlation coefficient.
All data points include error bars at the 1\,$\sigma$ level.
We should
emphasize that the measurements {\sl at fixed, say, velocity dispersion}
typically vary less than the trends with respect to the driving
parameter. As expected, the correlation coefficients of the trends
involving the intercepts are higher than those for the gradients, as
this measurement is less noisy\footnote{Differential measurements will
always carry larger uncertainties than integral ones.}.

\subsection{General trends}
\label{SSec:DiscussGeneral}

In this part of the discussion, we focus on the general trend,
shown by the grey data points. Of the six drivers, velocity dispersion
($\sigma$) appears to be the dominant one, with strongly correlated
trends in all observables, except, perhaps, $\nabla$[C/Fe].
This result is consistent with the analysis of \citet{Barone:18}, who
concluded that $g-i$ colour and total stellar metallicity
correlate stronger with the gravitational potential ($\Phi\propto\sigma^2$)
than with mass. This result is also in agreement with previous
work based on a larger sample of ETGs \citep[see, e.g.][]{Bernardi:03},
and with independent studies of samples extracted from the same
dataset \citep{Nic:17,Barone:18}. In the
following discussion, we will focus on this driver, with occasional
reference to the others.

[Z/H], as measured at R$_e$/4 ([Z/H]$_{e4}$), increases strongly with
$\sigma$. The radial gradients are overall substantially negative,
featuring a weak, negative trend
with respect to $\sigma$.  At the fiducial value
of velocity dispersion ($\sigma$=200\,\kms), the total metallicity is
unsurprisingly super-solar ([Z/H]=$+0.19\pm0.01$) with a strong
negative gradient ($\nabla$[Z/H]=$-0.31\pm0.02$). The slope of
[Z/H]$_{e4}$ is positive in all six drivers, although they are 
significant, in addition to $\sigma$, with stellar mass surface
density, stellar potential and $\sigma^2/R$. We note that the slope of
the metalliticy-velocity dispersion trend lies between $+0.46$ and
$+0.28$ (see Table~\ref{tab:fits}), values that are comparably
shallower than previous estimates, such as
$+0.58\pm0.05$ \citep{FLB:14} and $+0.65\pm 0.02$ \citep{Thomas:10}.
However, we should note that this work gives the metallicity at
R$_e$/4, whereas the quoted values correspond to an average
metallicity within the central regions of the galaxy. Also, as shown
in \citet[][see their table~5]{Harrison:11}, different studies find a
wide range of this slope, from $\sim+$0.18 to $\sim+$0.79.  These
variations are likely caused by different methods to derive the
stellar population parameters, as well as different selection criteria.
Our values of the metallicity-$\sigma$ slope fall
within the range reported by \citet{Harrison:11}

[Mg/Fe] features a strong positive slope regarding the central value
(i.e.  higher [Mg/Fe]$_{e4}$ with increasing $\sigma$), a well-known
correlation typically explained as a shorter duration of star
formation in more massive galaxies
\citep[see, e.g.][]{Trager:00,Thomas:05,IGDR:11}, as expected from
the delayed contribution of Fe-rich yields from type Ia supernovae
with respect to the $\alpha$-rich ejecta from type II.
In addition, we find  a strong
negative trend in the slope of $\nabla$[Mg/Fe] (i.e. more strongly decreasing
[Mg/Fe] outwards in more massive galaxies).  This trend is suggestive
of a more complex ex-situ formation scenario, where the stellar
component in the outer regions is populated by later stages of star
formation with higher chemical processing, therefore with a lower
[Mg/Fe]. The different slope in the trends of [Mg/Fe]$_{e4}$ and
$\nabla$[Mg/Fe] pose a significant caveat in the analysis, 
regarding the radial position at which [Mg/Fe] is
estimated. As we progress towards more massive galaxies, the
higher central [Mg/Fe] is compensated by a more negative gradient,
so that estimates of the [Mg/Fe] vs $\sigma$ slope will differ
greatly when evaluated at, say R$_e$/4 or 2R$_e$. Indeed,
\citet{Greene:15} found that, while in the galaxy centre [Mg/Fe]
increases with $\sigma$, this positive correlation disappears
as one moves towards larger galactocentric radii.
\cite{Parikh:19} analysed abundance pattern gradients as a function
of galaxy mass in a sample of SDSS/MaNGA galaxies. As shown in their
fig.~3, the positive correlation of [Mg/Fe] with mass in the galaxy
centre tends to disappear towards larger radii (beyond $\sim$R$_e$/2),
a qualitatively consistent result with \citet{Greene:15} and our work.

At the fiducial value of $\sigma$, the central value is
markedly super-solar ([Mg/Fe]$_{e4}$=$+0.19\pm0.01$\,dex) with a
rather shallow gradient ($\nabla$[Mg/Fe]=$-0.01\pm 0.01$). Note the
slope of [Mg/Fe] -- both the radial gradient and the central value --
do not correlate strongly with any of the other drivers, giving
more support to the idea that velocity dispersion, or equivalently,
the gravitational potential, is the major driver of the stellar
population content in ETGs.

[C/Fe], evaluated at R$_e$/4, shows a rather large amount of scatter, but, again, $\sigma$ is
the stronger driver, with a significant increasing trend of the
central value. This result agrees with previous work
\citep{Kelson:06,Graves:07,Schiavon:07,Smith:09,JTM:12}.
The slope of $\nabla$[C/Fe] is consistent with zero, in contrast with
the strong negative slope of $\nabla$[Mg/Fe] with respect to $\sigma$.
The trends in [C/Fe] roughly parallel those found for [Z/H]. At the
fiducial value of velocity dispersion, [C/Fe] is super-solar
([C/Fe]$_{e4}$=$+0.11\pm0.01$) with a negative radial gradient
($\nabla$[C/Fe]=$-0.10\pm 0.01$), once more a signature of a
substantially different population in the outer envelope.

Stellar age is also driven by velocity dispersion, with
$\delta_6\equiv\sigma^2/R$ becoming an equally strong driver, especially
if we consider that the correlation coefficient is higher in
$\delta_6$ (with respect to $\delta_1$) for the age at R$_{e4}$, and
also for the radial gradient of the age in the cluster subsample. Moreover,
note the slope of the total gravitational potential is 1/2 of the slope with respect
to $\sigma$.  The trend shows an increasing age with $\sigma$ as well
as a {\sl decreasing} radial gradient with $\sigma$, although we note
that at the fiducial value, this gradient is compatible with zero
($\nabla\log$Age=$-0.02\pm0.02$).  Given that most of our ETGs lie
below the fiducial value of $\sigma$, we conclude that at the low mass
end of the sample, the radial gradient of age is positive,
i.e. harbouring {\sl older} populations in the outer regions.

\subsection{Environment-related trends}
\label{SSec:DiscussEnviro}

We now turn our attention to differences in the observed trends between
cluster and field/group ETGs. Figs.~\ref{fig:Drivers_n} and \ref{fig:Drivers_e}
plot these two data sets independently, with cluster ETGs shown in
red and GAMA ETGs shown in blue. Tables~\ref{tab:fits} and \ref{tab:fits2}
also quantify the trends separately for each subsample.

It is quite remarkable to find relatively weak variations between a
cluster and a field/group environment. The stark contrast between
velocity dispersion and environment as drivers of the underlying
stellar populations has already been presented in previous
work \citep[see, e.g.][]{Rogers:10, Thomas:10, FLB:14, Greene:15,
Rosani:18}. In our data, most of the differences stay within the 
1-2\,$\sigma$ level, and need to rely on the linear correlation coefficient
to confirm these variations. The most conspicuous one is the slope
of [Z/H]$_{e4}$ with respect to velocity dispersion, with a steeper
slope in a cluster environment. This result would suggest that ETGs
at the massive end are more metal-rich in a cluster environment, a
result that could be explained by a higher chemical processing expected
in a system with more efficient star formation. This trend is followed,
although weakly, by [Mg/Fe]$_{e4}$ but it is intriguingly reversed
in the case of [C/Fe]$_{e4}$.

Although the metallicity gradient, $\nabla$[Z/H], is similar in both
types of environment, cluster ETGs show a strong negative change of
this gradient with increasing velocity dispersion, whereas the field
sample shows no measurable trend. This behaviour is consistent with a
scenario where field ETGs accrete more inhomogeneous material through
mergers, a possible sign of galactic conformity \citep{Weinmann:06},
whereby the stellar populations of galaxies within a group correlate
with those of the central galaxy. In a cluster environment, we
therefore expect that the properties of the merging progenitors were more
homogeneous with respect to their field counterparts.
The trends with [C/Fe] are similar to [Mg/Fe].  Note that the
correlation coefficient of the $\nabla$[Z/H] trends (bottom-left panel of
Fig.~\ref{fig:Drivers_n}) is more prominent in the cluster sample,
perhaps reflecting a more uniform star formation (and merger) history.

It is also worth mentioning the fiducial value of $\nabla$[Mg/Fe]
(at $\sigma$=200\,\kms), with a sizeable difference between cluster
and GAMA ETGs, the latter having a slightly negative radial gradient.
The fiducial $\nabla$[C/Fe] shows the opposite trend, with
cluster ETGs having a steeper, more negative radial gradient.
In contrast, the fiducial $\nabla$[Z/H] is the same in both subsamples.

Interestingly, by looking at the correlation coefficients, we note that
the other drivers sometimes feature stronger environment-related
differences. Such is especially the case with stellar mass.
As regards to age (i.e. differences in the time evolution of
star formation histories), no difference is found with respect
to environment. Only the  mass surface density (both stellar and $\sigma^2/R$) appear to show
a difference in the central value of age, with older ages
in cluster environments.
This result aligns with the proposal of \citet{Barone:18} of a
correlation between age and stellar mass density.

\section{Summary}
\label{Sec:Summary}

We study the radial gradients of early-type galaxies (ETGs) by use of
integral field unit data from the SAMI survey. Our working sample
comprises 522 visually classified ETGs located in the GAMA survey
(that maps field and group environments) as well as cluster
galaxies. This unique sample definition makes SAMI an ideal dataset to
explore environment-related mechanisms. In this case we focus on the
stellar population content, fitting radial gradients of total
metallicity, [Mg/Fe], [C/Fe] as well as stellar age.  A set of six
possible drivers are adopted (Table~\ref{tab:homog}, and
Fig.~\ref{fig:drivers}), and two main issues are sought: 1) to
identify the dominant driver of the radial gradients, and quantify the
trends, and 2) to determine the role of field/cluster environment in
the formation process of ETGs.

Our results (condensed in Figs.~\ref{fig:Drivers_n},  \ref{fig:Drivers_e},
and \ref{fig:Drivers_lAge}; and quantified in Tables~\ref{tab:fits},
and \ref{tab:fits2}) include a large amount of information that
should be used as constraints on numerical models of galaxy formation.
An incomplete, concise list of results follows:

\noindent
$\bullet$ The dominant driver controlling the stellar population properties of
ETGs is velocity dispersion ($\sigma$). Our work extends similar past claims
by looking in detail at a set of six different physical estimates
as possible drivers, finding that $\sigma$ is the one with strongly
correlated trends regarding radial gradients and central values of
the stellar population properties. We note that $\sigma$ is formally
equivalent to the total gravitational potential ($\Phi\propto\sigma^2$),
albeit with slopes differing by a factor of 1/2.

\noindent
$\bullet$ Surface mass density (regarding both total, $\delta_6$, and stellar, $\delta_4$),
also produces substantially strong correlations, especially with respect to
the radial gradient of stellar age.

\noindent
$\bullet$ Focusing on velocity dispersion (or gravitational potential) as the
main driver, we find a strong negative gradient of total metallicity
($\nabla$[Z/H]) with a weak dependence with respect to $\sigma$. [C/Fe]
appears to behave similarly to total metallicity but the correlation
is weaker. In contrast, the dependence of $\nabla$[Mg/Fe] on $\sigma$ is
quite steep and negative, so that the gradient is rather flat at the
massive end of this sample, turning to a strongly positive slope at the
low mass end. These trends suggest the merging
progenitors that contributed to populate the outer envelope of massive
ETGs during the `second stage' cannot be equivalent to low mass
galaxies at present. As regards to the central values (measured at R$_e$/4),
we find, unsurprisingly, substantial, positive slopes in all population indicators,
meaning that massive ETGs are older, more metal rich, and with higher abundance
ratios, a well-known result \citep[see, e.g.,][]{Renzini:06}.

\noindent
$\bullet$ Environment-related differences are subdominant, confirming previous
work in the literature. Our results quantify in detail the variations
in population gradients between a field/group and a cluster environment,
finding that in the central regions of galaxies (evaluating the
trends at R$_e$/4), cluster galaxies have more positively increasing
slopes of [Z/H] and [Mg/Fe] with $\sigma$, i.e. massive galaxies with the
strongest gravitational potential are more metal rich and [Mg/Fe] overabundant
in a cluster environment, with respect to the field. This trend intriguingly
reverses for [C/Fe], although the amount of scatter is rather high.
Environment-related differences in the trends regarding
radial gradients are harder to measure, but there is some evidence that
cluster ETGs have steeper (negative) slopes of the trend between $\nabla$[Z/H]
and $\sigma$ but no measurable difference in $\nabla$[Mg/Fe]. In contrast,
the value of $\nabla$[Mg/Fe] at fixed velocity dispersion appears shallower
in cluster galaxies.

\section*{Acknowledgements}
IF gratefully acknowledges support from the AAO through their
distinguished visitor programme, as well as funding from the Royal
Society. NS acknowledges support of a University of Sydney
Postdoctoral Research Fellowship. TB is supported by an Australian
Government Research Training Program Scholarship. JBH is supported by
an ARC Laureate Fellowship (FL140100278) that funds JvdS and an ARC
Federation Fellowship that funded the SAMI prototype.  SB acknowledges
the funding support from the Australian Research Council through a
Future Fellowship (FT140101166).  JJB acknowledges support of an
Australian Research Council Future Fellowship (FT180100231).  Support
for AMM is provided by NASA through Hubble Fellowship grant
\#HST-HF2-51377 awarded by the Space Telescope Science Institute, which
is operated by the Association of Universities for Research in
Astronomy, Inc., for NASA, under contract NAS5-26555.
MSO acknowledges the funding support from the Australian Research
Council through a Future Fellowship (FT140100255).
The SAMI Galaxy Survey is based on observations made at the
Anglo-Australian Telescope. The SAMI spectrograph was developed
jointly by the University of Sydney and the Australian Astronomical
Observatory. The SAMI input catalog is based on data from the Sloan
Digital Sky Survey, the GAMA Survey and the VST ATLAS Survey. The SAMI
Galaxy Survey is funded by the Australian Research Council Centre of
Excellence for All-sky Astrophysics (CAASTRO; grant CE110001020), and
other participating institutions.

\appendix

\setcounter{table}{0}
\renewcommand{\thetable}{A\arabic{table}}

\input SAMIGrad_tables.tex


\label{lastpage}

\end{document}

%% file: SAMIGrad_tables.tex
\begin{table*}
\caption{Radial gradients of chemical composition in SAMI ETGs (See text for details)\label{tab:fits}}
\begin{center}
\begin{tabular}{cc|ccc|ccc|}
\hline
 & & \multicolumn{3}{|c|}{$\nabla\pi$} & \multicolumn{3}{|c|}{[$\pi$]$_{e4}$}\\
\hline
$\pi$ & Env & Slope & Intercept & $r_{xy}$ & Slope & Intercept & $r_{xy}$\\
(1) & (2) & (3) & (4) & (5) & (6) & (7) & (8)\\
\hline
\multicolumn{8}{|c|}{Driver I: Velocity dispersion ($\log\sigma$)}\\
\hline
  & G & $+0.05\pm 0.23$  & $-0.31\pm 0.03$  & $-0.02\pm 0.06$ &  $+0.28\pm 0.11$  & $+0.17\pm 0.01$  & $+0.21\pm 0.05$\\
  & C & $-0.25\pm 0.16$  & $-0.30\pm 0.02$  & $-0.12\pm 0.06$ &  $+0.46\pm 0.08$  & $+0.20\pm 0.01$  & $+0.35\pm 0.04$\\
\multirow{-3}{*}{[Z/H]} & A & $-0.14\pm 0.13$  & $-0.31\pm 0.02$  & $-0.07\pm 0.04$ &  $+0.39\pm 0.07$  & $+0.19\pm 0.01$  & $+0.28\pm 0.03$\\
\hline
  & G & $-0.33\pm 0.14$  & $-0.05\pm 0.02$  & $-0.19\pm 0.06$ & $+0.15\pm 0.08$  & $+0.19\pm 0.01$  & $+0.10\pm 0.06$\\
  & C & $-0.28\pm 0.13$  & $+0.01\pm 0.02$  & $-0.14\pm 0.06$ & $+0.23\pm 0.05$  & $+0.19\pm 0.01$  & $+0.19\pm 0.04$\\
\multirow{-3}{*}{[Mg/Fe]} & A & $-0.30\pm 0.11$  & $-0.01\pm 0.01$  & $-0.16\pm 0.04$ & $+0.19\pm 0.05$  & $+0.19\pm 0.01$  & $+0.15\pm 0.03$\\
\hline
  & G & $+0.06\pm 0.16$  & $-0.07\pm 0.02$  & $-0.03\pm 0.05$ & $+0.26\pm 0.08$  & $+0.12\pm 0.01$  & $+0.23\pm 0.04$\\
  & C & $-0.06\pm 0.13$  & $-0.13\pm 0.01$  & $-0.05\pm 0.05$ & $+0.13\pm 0.06$  & $+0.11\pm 0.01$  & $+0.09\pm 0.05$\\
\multirow{-3}{*}{[C/Fe]} & A & $-0.03\pm 0.10$  & $-0.10\pm 0.01$  & $-0.04\pm 0.04$ & $+0.17\pm 0.04$  & $+0.11\pm 0.01$  & $+0.16\pm 0.03$\\
\hline
  & G & $-0.40\pm 0.24$  & $-0.02\pm 0.03$  & $-0.12\pm 0.05$ & $+0.48\pm 0.12$  & $+1.06\pm 0.02$  & $+0.25\pm 0.03$\\
  & C & $-0.33\pm 0.20$  & $-0.00\pm 0.03$  & $-0.14\pm 0.06$ & $+0.32\pm 0.11$  & $+1.11\pm 0.01$  & $+0.21\pm 0.04$\\
\multirow{-3}{*}{logAge} & A & $-0.39\pm 0.16$  & $-0.02\pm 0.02$  & $-0.13\pm 0.04$ & $+0.38\pm 0.09$  & $+1.09\pm 0.01$  & $+0.23\pm 0.03$\\
\hline
\multicolumn{8}{|c|}{Driver II: Stellar mass (log M$_s$)}\\
\hline
 & G & $+0.05\pm 0.07$  & $-0.27\pm 0.03$  & $+0.05\pm 0.06$ & $+0.06\pm 0.04$  & $+0.14\pm 0.02$  & $+0.14\pm 0.05$\\
 & C & $-0.11\pm 0.06$  & $-0.31\pm 0.02$  & $-0.17\pm 0.07$ & $+0.12\pm 0.03$  & $+0.21\pm 0.01$  & $+0.30\pm 0.05$\\
\multirow{-3}{*}{[Z/H]} & A & $-0.04\pm 0.04$  & $-0.29\pm 0.02$  & $-0.04\pm 0.04$ & $+0.10\pm 0.02$  & $+0.18\pm 0.01$  & $+0.21\pm 0.04$\\
\hline
 & G & $+0.01\pm 0.06$  & $-0.01\pm 0.02$  & $+0.04\pm 0.06$ & $-0.03\pm 0.03$  & $+0.18\pm 0.01$  & $-0.17\pm 0.06$\\
 & C & $-0.11\pm 0.06$  & $-0.01\pm 0.02$  & $-0.12\pm 0.06$ & $+0.05\pm 0.02$  & $+0.19\pm 0.01$  & $+0.11\pm 0.05$\\
\multirow{-3}{*}{[Mg/Fe]} & A & $-0.03\pm 0.03$  & $-0.01\pm 0.01$  & $-0.04\pm 0.04$ & $+0.01\pm 0.02$  & $+0.18\pm 0.01$  & $-0.03\pm 0.04$\\
\hline
  & G & $+0.03\pm 0.05$  & $-0.08\pm 0.03$  & $+0.01\pm 0.06$ & $+0.01\pm 0.03$  & $+0.09\pm 0.01$  & $+0.09\pm 0.05$\\
  & C & $+0.11\pm 0.05$  & $-0.12\pm 0.02$  & $+0.12\pm 0.06$ & $-0.04\pm 0.02$  & $+0.10\pm 0.01$  & $-0.12\pm 0.05$\\
\multirow{-3}{*}{[C/Fe]} & A & $+0.07\pm 0.03$  & $-0.10\pm 0.01$  & $+0.06\pm 0.04$ & $-0.03\pm 0.02$  & $+0.09\pm 0.01$  & $-0.01\pm 0.03$\\
\hline
  & G & $-0.11\pm 0.08$  & $-0.06\pm 0.03$  & $-0.16\pm 0.06$ & $+0.03\pm 0.05$  & $+1.06\pm 0.02$  & $+0.11\pm 0.04$\\
  & C & $-0.02\pm 0.07$  & $+0.00\pm 0.03$  & $-0.02\pm 0.06$ & $+0.01\pm 0.04$  & $+1.09\pm 0.02$  & $+0.05\pm 0.05$\\
\multirow{-3}{*}{logAge} & A & $-0.07\pm 0.04$  & $-0.03\pm 0.02$  & $-0.10\pm 0.03$ & $+0.02\pm 0.03$  & $+1.08\pm 0.01$  & $+0.07\pm 0.03$\\
\hline
\multicolumn{8}{|c|}{Driver III: Dynamical mass (log $M_d$)}\\
\hline
  & G & $+0.09\pm 0.06$  & $-0.29\pm 0.03$  & $+0.05\pm 0.06$ & $+0.02\pm 0.04$  & $+0.13\pm 0.01$  & $+0.07\pm 0.06$\\
  & C & $-0.06\pm 0.05$  & $-0.29\pm 0.02$  & $-0.08\pm 0.07$ & $+0.09\pm 0.02$  & $+0.17\pm 0.01$  & $+0.23\pm 0.05$\\
\multirow{-3}{*}{[Z/H]} & A & $-0.00\pm 0.03$  & $-0.28\pm 0.02$  & $-0.00\pm 0.04$ & $+0.05\pm 0.02$  & $+0.15\pm 0.01$  & $+0.14\pm 0.04$\\
\hline
 & G & $-0.06\pm 0.05$  & $-0.03\pm 0.02$  & $-0.13\pm 0.07$ & $+0.03\pm 0.02$  & $+0.19\pm 0.01$  & $+0.04\pm 0.05$\\
 & C & $-0.08\pm 0.04$  & $+0.03\pm 0.02$  & $-0.17\pm 0.05$ & $+0.08\pm 0.02$  & $+0.17\pm 0.01$  & $+0.24\pm 0.04$\\
\multirow{-3}{*}{[Mg/Fe]} & A & $-0.06\pm 0.03$  & $+0.01\pm 0.01$  & $-0.14\pm 0.04$ & $+0.05\pm 0.01$  & $+0.18\pm 0.01$  & $+0.14\pm 0.03$\\
\hline
  & G & $+0.05\pm 0.05$  & $-0.08\pm 0.02$  & $+0.06\pm 0.05$ & $+0.02\pm 0.02$  & $+0.10\pm 0.01$  & $+0.13\pm 0.05$\\
  & C & $-0.00\pm 0.04$  & $-0.13\pm 0.02$  & $-0.07\pm 0.05$ & $+0.01\pm 0.02$  & $+0.10\pm 0.01$  & $+0.05\pm 0.05$\\
\multirow{-3}{*}{[C/Fe]} & A & $+0.02\pm 0.03$  & $-0.11\pm 0.01$  & $-0.01\pm 0.04$ & $+0.01\pm 0.01$  & $+0.10\pm 0.01$  & $+0.09\pm 0.03$\\
\hline
  & G & $-0.11\pm 0.06$  & $-0.02\pm 0.03$  & $-0.15\pm 0.05$ & $+0.10\pm 0.04$  & $+1.05\pm 0.02$  & $+0.21\pm 0.04$\\
  & C & $-0.11\pm 0.06$  & $+0.01\pm 0.02$  & $-0.16\pm 0.06$ & $+0.07\pm 0.03$  & $+1.08\pm 0.01$  & $+0.17\pm 0.05$\\
\multirow{-3}{*}{logAge} & A & $-0.12\pm 0.04$  & $-0.00\pm 0.02$  & $-0.15\pm 0.04$ & $+0.08\pm 0.02$  & $+1.07\pm 0.01$  & $+0.19\pm 0.03$\\
\hline
\end{tabular}
\end{center}
\end{table*}


\begin{table*}
\caption{Radial gradients of chemical composition in SAMI ETGs (See text for details)\label{tab:fits2}}
\begin{center}
\begin{tabular}{cc|ccc|ccc|}
\hline
 & & \multicolumn{3}{|c|}{$\nabla\pi$} & \multicolumn{3}{|c|}{[$\pi$]$_{e4}$}\\
\hline
$\pi$ & Env & Slope & Intercept & $r_{xy}$ & Slope & Intercept & $r_{xy}$\\
(1) & (2) & (3) & (4) & (5) & (6) & (7) & (8)\\
\hline
\multicolumn{8}{|c|}{Driver IV: Surface stellar mass density ($\Sigma_s$)}\\
\hline
  & G & $-0.22\pm 0.09$  & $-0.52\pm 0.08$  & $-0.18\pm 0.05$ & $+0.22\pm 0.05$  & $+0.35\pm 0.05$  & $+0.26\pm 0.05$\\
  & C & $-0.04\pm 0.07$  & $-0.33\pm 0.07$  & $-0.03\pm 0.05$ & $+0.19\pm 0.03$  & $+0.34\pm 0.03$  & $+0.23\pm 0.04$\\
\multirow{-3}{*}{[Z/H]} & A & $-0.11\pm 0.05$  & $-0.41\pm 0.05$  & $-0.11\pm 0.04$ & $+0.20\pm 0.03$  & $+0.34\pm 0.03$  & $+0.25\pm 0.03$\\
\hline
  & G & $-0.05\pm 0.08$  & $-0.06\pm 0.08$  & $-0.03\pm 0.06$ & $-0.02\pm 0.04$  & $+0.16\pm 0.04$  & $+0.00\pm 0.05$\\
  & C & $-0.06\pm 0.07$  & $-0.02\pm 0.07$  & $-0.03\pm 0.05$ & $-0.01\pm 0.04$  & $+0.16\pm 0.04$  & $+0.02\pm 0.04$\\
\multirow{-3}{*}{[Mg/Fe]} & A & $-0.06\pm 0.06$  & $-0.05\pm 0.06$  & $-0.03\pm 0.04$ & $-0.01\pm 0.02$  & $+0.16\pm 0.02$  & $+0.01\pm 0.03$\\
\hline
  & G & $-0.04\pm 0.06$  & $-0.12\pm 0.07$  & $-0.03\pm 0.04$ & $+0.06\pm 0.04$  & $+0.15\pm 0.04$  & $+0.06\pm 0.05$\\
  & C & $-0.01\pm 0.07$  & $-0.14\pm 0.07$  & $-0.01\pm 0.05$ & $+0.11\pm 0.03$  & $+0.20\pm 0.03$  & $+0.20\pm 0.04$\\
\multirow{-3}{*}{[C/Fe]} & A & $-0.04\pm 0.05$  & $-0.15\pm 0.05$  & $-0.02\pm 0.04$ & $+0.09\pm 0.02$  & $+0.18\pm 0.02$  & $+0.13\pm 0.03$\\
\hline
  & G & $+0.07\pm 0.12$  & $+0.07\pm 0.12$  & $+0.08\pm 0.05$ & $-0.09\pm 0.05$  & $+0.96\pm 0.05$  & $-0.11\pm 0.04$\\
  & C & $-0.04\pm 0.10$  & $+0.00\pm 0.10$  & $-0.05\pm 0.05$ & $+0.07\pm 0.04$  & $+1.12\pm 0.05$  & $+0.12\pm 0.04$\\
\multirow{-3}{*}{logAge} & A & $+0.05\pm 0.08$  & $+0.06\pm 0.08$  & $+0.03\pm 0.04$ & $+0.00\pm 0.04$  & $+1.05\pm 0.04$  & $+0.00\pm 0.03$\\
\hline
\multicolumn{8}{|c|}{Driver V: Gravitational stellar potential ($\Phi_s$)}\\
\hline
 & G & $-0.11\pm 0.10$  & $-0.34\pm 0.05$  & $-0.12\pm 0.06$ & $+0.27\pm 0.07$  & $+0.23\pm 0.03$  & $+0.30\pm 0.06$\\
 & C & $-0.11\pm 0.09$  & $-0.35\pm 0.04$  & $-0.10\pm 0.07$ & $+0.24\pm 0.05$  & $+0.29\pm 0.03$  & $+0.36\pm 0.06$\\
\multirow{-3}{*}{[Z/H]} & A & $-0.12\pm 0.07$  & $-0.34\pm 0.03$  & $-0.11\pm 0.05$ & $+0.24\pm 0.04$  & $+0.26\pm 0.02$  & $+0.32\pm 0.04$\\
\hline
 & G & $-0.05\pm 0.11$  & $-0.04\pm 0.04$  & $-0.04\pm 0.07$ & $-0.02\pm 0.04$  & $+0.18\pm 0.02$  & $-0.05\pm 0.06$\\
 & C & $-0.04\pm 0.10$  & $-0.01\pm 0.04$  & $-0.01\pm 0.08$ & $+0.03\pm 0.04$  & $+0.20\pm 0.02$  & $+0.04\pm 0.05$\\
\multirow{-3}{*}{[Mg/Fe]} & A & $-0.05\pm 0.06$  & $-0.02\pm 0.02$  & $-0.02\pm 0.05$ & $+0.01\pm 0.03$  & $+0.19\pm 0.01$  & $-0.00\pm 0.04$\\
\hline
  & G & $-0.04\pm 0.11$  & $-0.09\pm 0.04$  & $-0.04\pm 0.06$ & $+0.08\pm 0.04$  & $+0.12\pm 0.02$  & $+0.15\pm 0.04$\\
  & C & $-0.03\pm 0.09$  & $-0.13\pm 0.03$  & $-0.08\pm 0.06$ & $+0.09\pm 0.03$  & $+0.13\pm 0.02$  & $+0.15\pm 0.05$\\
\multirow{-3}{*}{[C/Fe]} & A & $-0.05\pm 0.07$  & $-0.12\pm 0.03$  & $-0.06\pm 0.05$ & $+0.09\pm 0.03$  & $+0.13\pm 0.01$  & $+0.15\pm 0.03$\\
\hline
  & G & $+0.03\pm 0.13$  & $+0.03\pm 0.06$  & $+0.03\pm 0.06$ & $-0.02\pm 0.08$  & $+1.03\pm 0.04$  & $-0.01\pm 0.05$\\
  & C & $-0.14\pm 0.10$  & $-0.03\pm 0.05$  & $-0.12\pm 0.06$ & $+0.12\pm 0.06$  & $+1.12\pm 0.03$  & $+0.17\pm 0.04$\\
\multirow{-3}{*}{logAge} & A & $-0.07\pm 0.09$  & $-0.00\pm 0.04$  & $-0.04\pm 0.05$ & $+0.04\pm 0.04$  & $+1.08\pm 0.02$  & $+0.07\pm 0.03$\\
\hline
\multicolumn{8}{|c|}{Driver VI: Virial test ($\sigma^2/R$)}\\
\hline
 & G & $-0.09\pm 0.11$  & $-0.34\pm 0.04$  & $-0.10\pm 0.06$ & $+0.22\pm 0.06$  & $+0.21\pm 0.02$  & $+0.28\pm 0.05$\\
 & C & $-0.08\pm 0.10$  & $-0.30\pm 0.04$  & $-0.10\pm 0.07$ & $+0.20\pm 0.06$  & $+0.21\pm 0.02$  & $+0.23\pm 0.06$\\
\multirow{-3}{*}{[Z/H]} & A & $-0.09\pm 0.08$  & $-0.32\pm 0.03$  & $-0.10\pm 0.05$ & $+0.21\pm 0.04$  & $+0.21\pm 0.01$  & $+0.26\pm 0.04$\\
\hline
 & G & $-0.19\pm 0.10$  & $-0.08\pm 0.04$  & $-0.17\pm 0.07$ & $+0.10\pm 0.04$  & $+0.22\pm 0.01$  & $+0.23\pm 0.06$\\
 & C & $-0.08\pm 0.09$  & $+0.01\pm 0.03$  & $+0.00\pm 0.06$ & $+0.06\pm 0.05$  & $+0.18\pm 0.02$  & $+0.07\pm 0.05$\\
\multirow{-3}{*}{[Mg/Fe]} & A & $-0.11\pm 0.07$  & $-0.03\pm 0.02$  & $-0.08\pm 0.04$ & $+0.09\pm 0.03$  & $+0.20\pm 0.01$  & $+0.14\pm 0.04$\\
\hline
  & G & $-0.04\pm 0.09$  & $-0.10\pm 0.04$  & $-0.08\pm 0.06$ & $+0.12\pm 0.05$  & $+0.15\pm 0.02$  & $+0.20\pm 0.05$\\
  & C & $-0.08\pm 0.10$  & $-0.15\pm 0.03$  & $-0.05\pm 0.06$ & $+0.09\pm 0.04$  & $+0.11\pm 0.02$  & $+0.11\pm 0.05$\\
\multirow{-3}{*}{[C/Fe]} & A & $-0.06\pm 0.07$  & $-0.13\pm 0.02$  & $-0.06\pm 0.04$ & $+0.11\pm 0.03$  & $+0.13\pm 0.01$  & $+0.15\pm 0.04$\\
\hline
  & G & $-0.07\pm 0.14$  & $-0.04\pm 0.05$  & $-0.04\pm 0.06$ & $+0.21\pm 0.07$  & $+1.12\pm 0.03$  & $+0.24\pm 0.05$\\
  & C & $-0.27\pm 0.13$  & $-0.05\pm 0.04$  & $-0.24\pm 0.05$ & $+0.29\pm 0.07$  & $+1.15\pm 0.02$  & $+0.34\pm 0.04$\\
\multirow{-3}{*}{logAge} & A & $-0.17\pm 0.10$  & $-0.05\pm 0.03$  & $-0.13\pm 0.05$ & $+0.25\pm 0.06$  & $+1.13\pm 0.02$  & $+0.29\pm 0.03$\\
\hline
\end{tabular}
\end{center}
\end{table*}